\documentclass[a4paper,11pt]{article}

\usepackage{graphicx}
\usepackage{multicol} 
\usepackage{color}

\usepackage{amsmath}
\usepackage{amssymb}

\def\ie{{\it i.e. }}

\newcommand{\gev}{\,\mbox{GeV}}
\newcommand{\tev}{\,\mbox{TeV}}
\newcommand{\fb}{\,\mbox{fb}}
\newcommand{\mll}{M_{\ell\ell}}
\newcommand{\meff}{M_{\rm eff}}

\definecolor{rossoc}{cmyk}{0,0.5,1,0.2}
\definecolor{blu}{cmyk}{1,1,0,0.9}
\definecolor{blus}{cmyk}{1,1,0,0.9}
\definecolor{bluc}{cmyk}{1,1,0,0.1}
\definecolor{viola}{cmyk}{0,1,0,0.6}
\definecolor{verde}{cmyk}{0.92,0,0.59,0.25}
\definecolor{verdec}{cmyk}{0.92,0,0.59,0.15}
\definecolor{verdes}{cmyk}{0.92,0,0.59,0.4}

\definecolor{rosso}{cmyk}{0,1,1,0.4}
\definecolor{rossos}{cmyk}{0,1,1,0.9}
\definecolor{rossoc}{cmyk}{0,1,1,0.2}
\definecolor{blu}{cmyk}{1,1,0,0.3}
\definecolor{blus}{cmyk}{1,1,0,0.6}
\definecolor{bluc}{cmyk}{1,1,0,0.1}
\definecolor{verde}{cmyk}{0.92,0,0.59,0.25}
\definecolor{verdec}{cmyk}{0.92,0,0.59,0.15}
\definecolor{verdes}{cmyk}{0.92,0,0.59,0.4}
\definecolor{giallo}{cmyk}{0,0,1,0}
\definecolor{gialloverde}{cmyk}{0.44,0,0.74,0}
\definecolor{purple}{cmyk}{0.44,1,0.74,0}

\font\tenrsfs=rsfs10 at 12pt
\font\sevenrsfs=rsfs7
\font\fiversfs=rsfs5
\newfam\rsfsfam
\textfont\rsfsfam=\tenrsfs
\scriptfont\rsfsfam=\sevenrsfs
\scriptscriptfont\rsfsfam=\fiversfs
\def\mathscr#1{{\fam\rsfsfam\relax#1}}
\def\Lag{\mathscr{L}}

\newcommand{\fig}[1]{~\ref{fig:#1}}
\newcommand{\eq}[1]{~{\rm (\ref{eq:#1})}}

\newcommand{\GeV}{\,{\rm GeV}}
\newcommand{\TeV}{\,{\rm TeV}}
\newcommand{\MeV}{\,\hbox{\rm MeV}}

\def\circa#1{\,\raise.3ex\hbox{$#1$\kern-.75em\lower1ex\hbox{$\sim$}}\,}

\newcommand{\beq}{\begin{equation}}
\newcommand{\eeq}{\end{equation}}

\def\circa#1{\,\raise.3ex\hbox{$#1$\kern-.75em\lower1ex\hbox{$\sim$}}\,}
\makeatletter

\def\art{\@ifnextchar[{\eart}{\oart}}
\def\eart[#1]#2#3#4#5#6{{\rm #2}, {\em #3 \bf #4} {\rm (#6) #5} ({\em #1})}
\def\hepart[#1]#2{{\rm #2, \em#1}}
\newcommand{\oart}[5]{{\rm #1}, {\em #2 \bf #3} {\rm (#5) #4}}

\newcounter{alphaequation}[equation]
\def\thealphaequation{\theequation\hbox to
0.6em{\hfil\alph{alphaequation}\hfil}}
\def\eqnsystem#1{
\def\@eqnnum{{\rm (\thealphaequation)}}
\def\@@eqncr{\let\@tempa\relax \ifcase\@eqcnt \def\@tempa{& & &} \or
  \def\@tempa{& &}\or \def\@tempa{&}\fi\@tempa
  \if@eqnsw\@eqnnum\refstepcounter{alphaequation}\fi
\global\@eqnswtrue\global\@eqcnt=0\cr}
\refstepcounter{equation} \let\@currentlabel\theequation \def\@tempb{#1}
\ifx\@tempb\empty\else\label{#1}\fi
\refstepcounter{alphaequation}
\let\@currentlabel\thealphaequation
\global\@eqnswtrue\global\@eqcnt=0 \tabskip\@centering\let\\=\@eqncr
$$\halign to \displaywidth\bgroup \@eqnsel\hskip\@centering
$\displaystyle\tabskip\z@{##}$&\global\@eqcnt\@ne
\hskip2\arraycolsep\hfil${##}$\hfil& \global\@eqcnt\tw@\hskip2\arraycolsep
$\displaystyle\tabskip\z@{##}$\hfil
\tabskip\@centering&\llap{##}\tabskip\z@\cr}
\def\endeqnsystem{\@@eqncr\egroup$$\global\@ignoretrue} \makeatother

\font\tenrsfs=rsfs10
\font\sevenrsfs=rsfs7
\font\fiversfs=rsfs5
\newfam\rsfsfam
\textfont\rsfsfam=\tenrsfs
\scriptfont\rsfsfam=\sevenrsfs
\scriptscriptfont\rsfsfam=\fiversfs
\def\mathscr#1{{\fam\rsfsfam\relax#1}}
\def\Lag{\mathscr{L}}

\def\Emisst{E_T\hspace{-2.6ex}/\hspace{2ex}}

\def\beq{\begin{equation}}
\def\eeq{\end{equation}}
\def\bea{\begin{eqnarray}}
\def\eea{\end{eqnarray}}

\begin{document}
\centerline{hep-ph/0408320 \hfill
IFUP--TH/2004-18 \hfill   CERN-TH/2004-160}

\vspace{1cm}
\centerline{\LARGE\bf \color{rossos}Graviton collider effects 
in one}\vspace{3mm}
\centerline{\LARGE\bf  \color{rossos}and more large extra dimensions}
\vspace{0.3cm}
\bigskip\bigskip

\begin{center}
{\large\bf Gian F. Giudice, Tilman Plehn}  

\medskip

{\em Theoretical Physics Division,  CERN, CH-1211,  Geneva 23, Switzerland}

\bigskip

{\large and {\large\bf Alessandro Strumia}}  

\medskip

{\it Dipartimento di Fisica dell'Universit{\`a} di Pisa and INFN,  Italy}
\end{center}

\bigskip\bigskip\color{blus}
\centerline{\large\bf Abstract}\begin{quote}\large
Astrophysical bounds severely limit the possibility of observing collider
signals of gravity with less than 3 flat extra dimensions.
However, small distortions of the compactified space can lift the masses
of the lightest 
graviton excitations, evading astrophysical bounds without affecting
collider signals of quantum gravity.
Following this procedure we reconsider theories with one large extra dimension.
A slight space warping gives a model which is safe in the infrared 
against astrophysical and observational bounds,
and which has the ultraviolet properties of gravity 
with a single flat extra dimension.
We extend collider studies to the case of one extra dimension, pointing 
out its peculiarities.
Finally, for a generic number of extra dimensions, we compare
different channels in LHC searches for quantum gravity, introducing
an ultraviolet cutoff as an additional parameter besides the Planck mass.
 \end{quote}\color{black}
\noindent


\section{Introduction}
LHC experiments will hopefully allow us to understand
why the electroweak scale is much smaller than the Planck scale.
This hierarchy can arise  from a fundamental $(4+\delta )$-dimensional Planck mass close
to the electroweak scale if gravity propagates
in $\delta$ extra dimensions compactified on a large volume~\cite{add}. 
This hypothesis has important phenomenological
consequences, since high-energy colliders could probe the dynamics of
gravity in its quantum and semi-classical regimes.
 
The prospects for observing graviton-induced processes at future collider
experiments, in the case of 2 and 3 flat extra dimensions, are very much 
limited by the present astrophysical bounds~\cite{astro}. Graviton emissions
in supernov\ae{} and neutron stars set a limit on the $D$-dimensional 
($D=4+\delta$)
gravitational mass scale $M_D$ of about 40~TeV for $\delta=2$ and
3~TeV for $\delta=3$, while the limits for larger values of $\delta$
are below the TeV scale~\cite{astro}. These bounds apparently rule 
out the possibility
of testing the theory at the LHC for $\delta=2$, and severely restrict
the available parameter space for $\delta=3$. However, it is well known
that, while astrophysics probe only the infrared
end of the Kaluza-Klein (KK) spectrum of gravitons, high-energy experiments
are mainly sensitive to the ultraviolet side. Therefore, one can conceive
non-trivial compactification spaces for which the large volume determines
the hierarchy between the multi-dimensional and 4-dimensional Planck 
scales ($V\sim M_{\rm Pl}^2/M_D^{2+\delta}$), 
and in which the mass of lightest KK excitation
is not given by the inverse radius $1/R\sim V^{-1/\delta}$, 
but by a new intrinsic mass
$\mu$. If $\mu\circa{>}50$~MeV gravitons cannot be produced 
in astrophysical environments, which therefore give no bounds on the scale
$M_D$. Whenever $\mu$ is smaller than 
the characteristic energy resolution of high-energy
experiments, collider predictions are not affected by its presence.

Therefore, LHC quantum-gravity searches for $\delta =2$ and 3 are still
viable, in spite of the strong astrophysical bounds\footnote{Incidentally, 
we should add that, on the contrary, the astrophysical bounds
directly rule out the possibility of observing graviton-induced signals
in experiments testing gravity in the micron-to-millimeter region.}.    
Undoubtedly, 
it is quite
disturbing that the new geometrical mass scale $\mu$ has a size which
is unrelated to the other scales $V^{-1/\delta}$ or $M_D$, and (unless
a satisfying  justification is found) it appears
that its only
purpose is to make the theory
evade the astrophysical bounds. Nevertheless, because of 
the great interest in investigating quantum gravity at colliders, 
studies of $\delta =2$ and 3 are still actively pursued.

\medskip

On the other hand, the case $\delta=1$ has been discarded immediately.
This is because Newtonian gravity would be modified at the macroscopic scale
$M_{\rm Pl}^2/(2\pi M_5^3)=(\TeV /M_5)^3$~AU, and this is excluded by
astronomical observations.
We have just seen that the large distance behavior of the theory can be mended by introducing
an appropriate KK mass gap $\mu$ that eliminates 
graviton effects in astrophysics.
However, this remedy does not appear immediately
applicable to the case $\delta =1$. The reason is that a one-dimensional
compact manifold cannot have a non-trivial shape
that makes the lightest KK heavier than $1/R$.
 

In this paper we show that it is indeed 
possible to distort the ADD model~\cite{add}
with $\delta=1$ in such a way that the infrared regime coincides with
Newtonian gravity up to a distance determined by the inverse of the
KK mass gap
$\mu$, chosen to avoid any bound from astrophysics or large-distance
observations. The ultraviolet behavior of ADD with
$\delta=1$ remains unmodified. Although the value of $\mu$ is chosen {\it
ad hoc} for phenomenological reasons, this is completely analogous to
the procedure required for the cases $\delta =2$ and 3. Therefore, 
LHC searches should not dismiss
the case $\delta =1$ which, as we  show in this paper, present
interesting peculiarities.

The crucial point is that collider searches provide a test of the theory
which is very robust and independent of the details of the compactification
geometry. This can be simply understood by noting that the high-energy
predictions are valid in the limit $V\to \infty$ (in which ordinary
gravity decouples, $M_{\rm Pl}\to \infty$) and $\mu \to 0$ (in which 
the KK mass gap is neglected). High-energy collisions effectively see
a $D$-dimensional infinite, non-compact and flat space. On the other
hand, astrophysical bounds crucially depend on the details of the geometry
and of the compactification, and are most sensitive to $\mu$ and $V$.

\bigskip

A second aspect of extra-dimensions  discussed in this paper
is the comparison between searches for graviton emission and contact 
interactions at the LHC. 
Graviton emission  can be
consistently computed by linearizing Einstein gravity, as long as the
relevant energy is sufficiently smaller than $M_D$. For an $e^+e^-$ collider,
this condition has to be applied simply to the value of $\sqrt{s}$
at which the collider operates.
But at the LHC,
where parton collisions can occur at very different center-of-mass
energies, some care is required~\cite{noi} to define the validity region
of perturbation theory. 

On the other hand, unlike graviton emission,
the exchange of virtual gravitons
is dominated by the ultraviolet region and it is 
therefore not calculable in the effective Einstein theory, without
specific knowledge of its short-distance completion. In ref.~\cite{stru}
we introduced a prescription to parametrize this ignorance in terms of an
ultraviolet cutoff $\Lambda$, allowing for a possible comparison
between graviton emission, tree-level graviton exchange, and graviton
loops. In particular, it was shown that, for small values of $\Lambda$,
direct graviton production at LEP provides the strongest constraint
on $M_D$ while, for moderate and large $\Lambda$, the most stringent
constraint arises from LEP limits on a dimension-6 axial-axial
effective operator which is induced by loops of gravitons. In this
paper, we extend the analysis to searches at the LHC, including
also the case $\delta =1$.

\medskip

This paper is organized as follows.
In section~\ref{seccaz} we show how it is possible to distort one large extra dimension
avoiding unwanted astrophysical effects,
without affecting the signals at high-energy colliders.
In section~\ref{5d} we compute these signals, which are
somewhat different than in the cases $\delta >1$.
In section~\ref{LEP} we study present data, 
obtaining a bound $M_D\circa{>}2\TeV$ for $\delta=1$.
In section~\ref{LHC} we consider LHC signals and compare the
discovery reach of the different channels, for a generic number of
extra dimensions $\delta$.
Appendix~\ref{cross} contains explicit formul{\ae} for
graviton effects.

\section{IR modifications of gravitons in large extra dimensions}\label{seccaz}

In this section we present a distorted version of the ADD model~\cite{add}
with one extra dimension ($\delta =1$),
which has the same properties of the original model in the short-distance
region (probed by colliders), but satisfies observational and
astrophysical limits in the large-distance regime.
The distortion introduced to forbid the unwanted
 low-energy effects
corresponds to the RS1 model~\cite{rs1}, in
the limit in which both the compactification and the AdS radii are
large with respect to the inverse of the fundamental gravity scale $M_5$.
Pictorially, we consider a slightly warped but long extra dimension,
that results in a moderately large total warp factor.
The hierarchy between the Fermi and the Planck scale is generated by two
factors:  the large extra dimension, and warping.


We start from the RS1 model, choosing the coordinates such that the
visible brane is located at $y=0$ and the Planck brane at $y=\pi R$ and
the line element is
\beq
ds^2=e^{2\sigma(y)}\eta_{\mu\nu}dx^\mu dx^\nu +dy^2,~~~~\sigma(y)\equiv
\mu |y|
\eeq
with  $0\le |y|\le \pi R$. The 4-dimensional Planck mass is given 
by\footnote{Throughout the paper $M_{\rm Pl}=2.4\times 10^{18}$~GeV
is the reduced 4-dimensional Planck mass. We define ${\bar M}_5$
to be the reduced 5-dimensional Planck mass which, following the
notation of ref.~\cite{noi}, is related to the Planck mass by
$M_5=(2\pi)^{1/3} {\bar M}_5$.}
\beq\label{eq:MPl}
M_{\rm Pl}^2 =\frac{{\bar M}_5^3}{\mu}\left( e^{2\mu R\pi}-1\right).
\eeq
In the limit of small $\mu$ ($\mu \ll R^{-1}$), eq.\eq{MPl} reduces to the usual
flat-space relation  $M_{\rm Pl}^2 ={\bar M}_5^3 V$, where the ``volume'' 
of the circle
is $V=2\pi R$. However, if $\mu$ is smaller than the inverse radius,
it cannot really affect the KK mass gap. So we are interested in the
case in which $\mu$ is larger than $R^{-1}$ (enough to damp processes in
astrophysical environments), but both are much smaller than ${\bar M}_5$.

We expand the metric fluctuation as
\beq
h_{\mu\nu}(x,y)=\sum_{n=0}^{\infty}\frac{h_{\mu\nu}^{(n)}(x)}{\sqrt{2\pi R}}
\phi^{(n)}(y),
\eeq
where $h_{\mu\nu}^{(n)}(x)$ has the canonical propagator as defined in 
ref.~\cite{noi}.
The KK masses are obtained by computing the eigenvalues of the
operator $p_y^2$, corresponding to the square of the momentum along the 
fifth direction in warped space,
\beq  
\left[ e^{-2\sigma}\frac{d}{dy}\left( e^{4\sigma}\frac{d}{dy}\right)
+m_n^2 \right]\phi^{(n)}(y)=0.
\label{masse}
\eeq
The graviton eigenfunctions are even, $\phi^{(n)}(-y)= \phi^{(n)}(y)$,
satisfy the boundary conditions $d\phi^{(n)}/dy =0$ at $y=0$
and at $y=\pi R$, and the orthonormality relation
$\int_{-\pi R}^{\pi R}dy~e^{2\sigma}\phi^{(n)}(y)\phi^{(m)}(y)=2\pi R\delta_{n m}$.

The solutions to eq.~(\ref{masse}) are
\beq
\label{ripaz}
\phi^{(n)}(y)=\frac{z_n^2}{N_n}\left[ J_2(z_n)+c_nY_2(z_n)\right] ,
~~~~z_n\equiv \frac{m_n}{\mu}e^{-\sigma(y)},
\eeq
where $J_\nu$ and $Y_\nu$ are Bessel functions\footnote{We list some 
properties 
of the Bessel functions we used to obtain our results:
$$J_{\nu+1}(x) Y_{\nu}(x)-J_{\nu}(x)Y_{\nu+1}(x) = 2/(\pi x),~~~
J_{\nu+2}(x) Y_{\nu}(x)-J_{\nu}(x)Y_{\nu+2}(x) = 4(\nu+1)/(\pi x^2),$$
$$J_1(x)\stackrel{x\to \infty}{\simeq}
\sqrt{2/(\pi x)}\cos ( x-3\pi/4),~~~
Y_1(x)\stackrel{x\to \infty}{\simeq}
\sqrt{2/(\pi x)}\sin ( x-3\pi/4),$$
$$J_1(x)\stackrel{x\to 0}{\simeq}x/2 ,~~~
Y_1(x)\stackrel{x\to 0}{\simeq} -2/(\pi x).$$}.
The integration constants $N_n$ and $c_n$ are determined by the 
orthonormality condition and by the first 
boundary condition
\beq
N_n^2=\frac{2a_n^2}{\mu R \pi^3}\left[ \frac{1}{Y_1^2(a_n)}-
\frac{1}{Y_1^2(b_n)}\right],~~~~
c_n=-\frac{J_1(a_n)}{Y_1(a_n)},
\eeq
\beq
a_n\equiv \frac{m_n}{\mu},~~~~b_n\equiv a_n e^{-\mu R\pi}.
\eeq
The second boundary condition determines the masses $m_n$ from the equation
\beq
J_1(a_n)Y_1(b_n)=J_1(b_n)Y_1(a_n).
\label{ahmasse}
\eeq

\begin{figure}
$$\includegraphics[width=17cm]{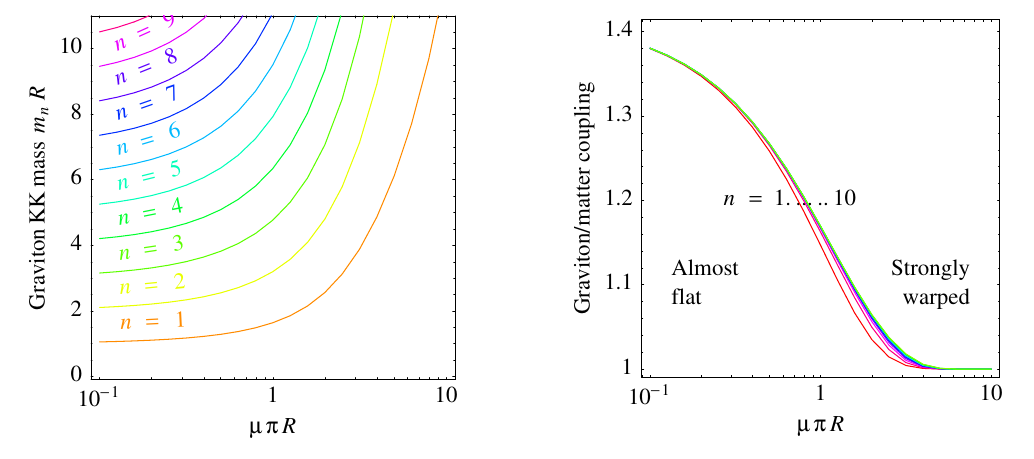}$$
\caption{\label{fig:Lambdapi}\em Masses and couplings to matter
of the first 10 Kaluza-Klein
excitations of gravitons,
as functions of the warp factor $\mu\pi R$.
The masses $m_n$ are given in units of $1/R$;
couplings in units of $e^{\mu\pi R}/M_{\rm Pl}$.}
\end{figure}

The interactions of massive gravitons are described by the Lagrangian
density
\beq
\Lag=-\frac{1}{{\bar M}_5^{3/2}}\int dy~ h_{\mu \nu}(x,y) T^{\mu\nu}(x)
\delta (y)=
-T^{\mu \nu} \sum_{n=1}^{\infty}\frac{1}{\Lambda_n} h_{\mu \nu}^{(n)}
(x),
\label{ahlamm}
\eeq
with
\beq
\Lambda_n  \equiv \frac{{\bar M}_5^{3/2}\sqrt{2\pi R}}{\phi^{(n)}(0)}
\eeq
Using eqs.~(\ref{eq:MPl}) and (\ref{ripaz}), the graviton interaction
scale can be written as
\beq
\Lambda_n =
M_{\rm Pl}\sqrt{\frac{1-J_1^2(a_n)/J_1^2(b_n)}{e^{2\mu R\pi}-1}},
\label{lamm}
\eeq
and it is shown in fig.~\ref{fig:Lambdapi}, where we plot
the masses and the matter couplings of the first 10 KK excitations of
the graviton as functions of the warp factor.
We see that warping removes the lightest KK's with mass $\sim1/R$ from
the spectrum.

The expressions written so far are valid for any value of $\mu$. 
First consider the limit of small AdS inverse radius $\mu \ll R^{-1}$,
in which eq.~(\ref{ahmasse}), determining the mass eigenvalues,
becomes $\sin (m_nR\pi)=0$. This leads to
the 
equally-spaced KK masses $m_n=n/R$, characteristic of flat space. 
In the same limit, eq.~(\ref{lamm}) becomes $\Lambda_n =
M_{\rm Pl}/\sqrt{2}$, recovering the ADD result~\footnote{We are summing
only over positive KK indices $n$, and this explains the extra $\sqrt{2}$
factor.}.

However, we are interested here in the case
$\mu >R^{-1}$. Then eq.~(\ref{ahmasse}) is approximately equal 
to
$J_1(a_n)=0$, whose solutions are
$m_n=x_n \mu$ with \beq
x_1\approx 3.8,\qquad
x_2\approx 7.0,\qquad
x_n=\beta_n-\frac{3}{8\beta_n}
+\frac{3}{128\beta_n^3}-\frac{1179}{5120\beta_n^5}+\dots ,~~~
\beta_n\equiv \left( n+\frac{1}{4}\right)\pi .
\eeq
Notice that at large $n$ KK states are again equally spaced.

For a given KK mass gap $m_{\rm GAP} = m_1$,
we obtain 
\beq
\mu =\left(\frac{m_{\rm GAP}}{50~\MeV}\right) ~13~\MeV,
\eeq
\beq
R=\frac{1}{2\pi\mu}\ln \frac{\mu M_{\rm Pl}^2}{{\bar M}_5^3} =
\left(\frac{50~\MeV} {m_{\rm GAP}}\right) 
\left[ 1.9 +1.2\times 10^{-2}\ln\left(\frac{m_{\rm GAP}}{50~\MeV}\right)
\left(\frac{\TeV}{{\bar M}_5}\right)^3\right] \MeV^{-1}.
\eeq
Therefore, $\mu$ is the parameter which controls the mass gap and its
minimum value is determined by the astrophysical bounds. The KK mass gap
is large enough to avoid clash with observations,
but still small enough to be unessential in the
ultraviolet region where high-energy experiments are performed.
The warp factor $\exp(\mu R\pi)\sim M_{\rm Pl} \mu^{1/2}/\bar{M}_5^{3/2}$ is large, 
but much smaller than in RS1, where $\exp(\mu R\pi)\sim M_{\rm Pl} /\bar{M}_5$.

For $\mu >R^{-1}$, the graviton interaction scale becomes 
$\Lambda_n=M_{\rm Pl} \exp(-\mu R\pi)$, which can be expressed as
\beq
\Lambda_n  \simeq \frac{{\bar M}_5^{3/2}}{\mu^{1/2}}=
\left(\frac{{\bar M}_5}{\TeV}\right)^{3/2}
\left(\frac{50~\MeV} {m_{\rm GAP}}\right)^{1/2} 2.8\times 10^5~\GeV .
\eeq 
Notice that both in the limit of small or large $\mu R$, the
interaction scale $\Lambda_n$ is independent of $n$ and
all KK gravitons have a universal coupling. This is not true in the
intermediate region, as illustrated in fig.~\ref{fig:Lambdapi}.

If the energy resolution in a collider experiment is larger than 
the graviton KK mass splitting ($\sim \pi \mu$),
it is more appropriate to consider inclusive processes, in which we sum
over the individual contributions of the KK modes. 
Using a continuous, instead of a discrete, description of the KK spectrum,
the density of states in the original ADD model with $\delta =1$, expressed
in terms of the KK masses $m$, is
\beq
dn=R~dm=\frac{M_{\rm Pl}^2}{2\pi{\bar M} _5^3}~dm .
\eeq
In the case of the distorted ADD model under consideration, we find
\beq
dn=\left( \frac{dx_n}{dn}\right)^{-1} \frac{dm}{\mu},~~~~
\frac{dx_n}{dn}=\pi \left( 1+\frac{3}{8\beta_n^2}+\dots \right) .
\eeq
We write the $n$-th KK graviton production cross section as $\hat{\sigma}_n
/\Lambda_n^2$, where we have factored out the interaction mass scale
defined in eq.~(\ref{ahlamm}). We obtain that the inclusive cross section,
both in the case of undistorted and distorted ADD is given by
\beq
\sum_{n=1}^\infty \frac{\hat{\sigma}_n}{\Lambda_n^2}=\frac{1}{\pi 
{\bar M}_5^3}\int_0^\infty dm
~\hat{\sigma}(m) .
\eeq
Therefore, although in the distorted model the state distribution 
is less dense, the KK modes interact more strongly than in the undistorted
model, in such a way that the two effects exactly compensate in inclusive
cross sections. The result is easily understood: if a collider experiment
is not sensitive to the aspects of the discretization ({\it i.e.} if
the energy resolution is larger than both $\mu$ and $1/R$), the two models are
completely indistinguishable. The two models differ only in the far infrared,
where astrophysical probes are important.

The equivalence between the two models holds also for effective interactions
mediated by gravitons, as long the typical energy of the relevant physical
process is larger than the KK mass gap (and we are allowed to go from 
the discrete to the continuum). For instance, graviton tree-level exchange
in the $s$-channel leads to the scattering amplitude
\beq\label{eq:T}
{\cal A} ={\cal S}(s) {\cal T},~~~~{\cal T}\equiv  T_{\mu\nu}T^{\mu\nu}-
\frac{1}{3}T^\mu_\mu T^\nu_\nu ,
\eeq
\beq
{\cal S}(s)=\sum_{n=1}^\infty \frac{1}{\Lambda_n^2}
\frac{1}{s-m_n^2+i\varepsilon}=\frac{1}{\pi{\bar M} _5^3}
\int_0^\infty \frac{dm}{s-m^2+i\varepsilon}.
\eeq
which agrees with the usual definition in $D$ dimensions~\cite{noi}
\beq {\cal S}(s)=\frac{1}{M _D^{2+\delta}}\label{spuf}
\int_{-\infty}^{+\infty} ~\frac{d^\delta q}{s-q^2+i\varepsilon}.\eeq

\subsection*{Extensions to $\delta >$ 1}

The deformation of the $\delta =1$ ADD model presented above
can be generalized to $\delta >1$. Particularly interesting is the case
$\delta =2$, where a distortion is needed to confront the strong
astrophysical bounds.

The simplest generalization can be obtained by following the construction
described in ref.~\cite{inter}. Let us consider $\delta$ (with $\delta >1$)
non-parallel $(2+\delta)$ branes which live in a $(4+\delta )$ dimensional 
space. Each brane has co-dimension 1, and therefore for each one we can
construct an RS model with an AdS mass $\mu$ which acts as a mass gap.
The intersection of the branes, which has dimension $(2+\delta)-(\delta -1)=3$,
is taken to be the ordinary space where the SM fields are confined.
In this case, one can explicitly compute the mass spectrum in presence
of the distortion parameter $\mu$. Another possibility is to start
from the 6-dimensional models with gravity localized on string-like 
defects~\cite{shap}.

For $\delta >1$, one can use an alternative approach and consider
non-trivial compactification spaces, as the compact hyperbolic
manifold, considered in ref.~\cite{rus}. 
In this case, the volume-radius ratio
$V/R^\delta$ is a coefficient determined by the topology, which can be 
much larger than one. By making an appropriate choice of the largest
distance inside the manifold, it is possible to arbitrarily increase
the KK mass gap, for a fixed value of the volume $V$. 

A different example~\cite{keit} 
is given by a general 2-dimensional torus determined by its
volume $V$, by the ratio between the two radii $r$ (choosing $r\le 1$), and
by the shift angle $\theta$ (with $0< \theta \le \pi/2$). The KK
mass spectrum is given by
\beq
m_n^2=\frac{4\pi^2}{V\sin\theta}\left( n_1^2r+\frac{n_2^2}{r}-2n_1n_2
\cos\theta \right) ,
\label{spec}
\eeq
where $n_{1,2}\in {\bf Z}$. For $\theta =\pi /2$, the
mass square of the
first KK mode is $4\pi^2r/V$, and therefore departing from the standard
value $r=1$ can only reduce the mass gap. When $\theta \to 0$, it is
always possible to find values of $n_{1,2}$ such that the term inside
brackets in eq.~(\ref{spec}) vanishes, as long as $r$ is rational. Therefore,
for rational $r$, 
the mass gap goes to zero as $\theta \to 0$. According to ref.~\cite{keit},
an interesting twist arises
when $r$ is an algebraic irrational number. In this case, 
for $\theta \to 0$, the mass gap determined by eq.~(\ref{spec}) 
goes to a constant, which can then be chosen arbitrarily large, for any
fixed value of $V$. For a careful choice of parameters, one can appropriately
tune the mass gap to the preferred value of $\mu$.

\section{Graviton signals in one large extra dimension}\label{5d}
The distortion introduced in the previous section
to cure the low-energy phenomenology of gravitons in large extra dimensions
does not affect their high-energy signals.
Therefore we can study them by setting $\delta=1$ in the expressions,
valid for a generic number of extra dimensions,
for graviton-emission processes~\cite{noi} and graviton-loop induced
interactions~\cite{stru}. 
An interesting peculiarity of $\delta =1$ occurs in processes generated
by tree-level graviton exchange.
Contrary to
the cases with more extra dimensions, the integral in eq.~(\ref{spuf})
is not divergent in the ultraviolet. 
For $\delta=1$ the integral is dominated by KK masses of the order of $\sqrt{s}$.
Introducing an ultraviolet cutoff
$\Lambda$, 
which corresponds to the largest KK graviton
mass, we obtain
\beq\label{eq:S(s)}
{\cal S}(s)=\frac{2}{M_5^3\sqrt{s}}~{\rm arctanh}\frac{\Lambda}
{\sqrt{s}}
\simeq \frac{1}{M_5^3} \left( \frac{-i\pi}{\sqrt{s}}+\frac{2}{ 
\Lambda}\right) .
\eeq

In the limit $\Lambda\to \infty$, 
${\cal S}(s)$ is actually purely imaginary, 
and therefore there is no interference with Standard Model contributions
to the same physical process. 
The imaginary part in ${\cal S}(s)$ corresponds to the exchange of a 
resonant graviton, as can be seen by giving a small width $\Gamma$ to the
KK excitations.
The physical value of the width 
can be smaller or larger than the separation
between consecutive KK, giving rise to a series of separated or of overlapped resonances.
Studying experiments that do not have enough energy resolution to see the difference,
it is convenient to take the simpler limit $\Gamma\to 0$, where
\beq
\lim_{\Gamma\to 0}\frac{1}{s-m^2 + i m \Gamma } = P\left(
\frac{1}{s-m^2}\right) - i \pi \delta(s-m^2) .
\eeq
Physically this means that at any $s$ the dominant effect is 
on-shell production
of the KK mode with mass $m=\sqrt{s}$.
This is why graviton exchange does not interfere with the SM
amplitude, as mathematically accounted by the $i$ factor in eq.\eq{S(s)}.
The unknown physics that acts as cut-off of quantum gravity
can give the uncontrollable extra contribution suppressed by $\Lambda$ in eq.\eq{S(s)},
which can be real and change the above conclusion.

\medskip

Effects mediated by virtual graviton exchange in the $t$-channel are
qualitatively different, because now
gravitons cannot be resonantly produced.
Indeed, inserting $t<0$ in the expression for ${\cal S}$, eq.\eq{S(s)},
we get that the $t$-channel
contribution is purely real, ${\cal S}(t)<0$.
Therefore the $t$-channel contribution can interfere with the SM amplitude.
When the SM cross section is dominant, 
$t$-channel graviton exchange gives the only significant correction.
Consequently particular attention should be paid to 
elastic processes, like Bhabha
scattering at LEP, electron-jet final states at HERA, and dijet production
at Tevatron and LHC.

%


\medskip

The dependence of ${\cal S}$ on the kinematical variables of the process
signals a non-local effect. 
This can be interpreted as the result of
a large anomalous dimension of the local
dimension-8 operator $\cal T$ corresponding to the linear (instead of 
logarithmic) running in 5 dimensions. Effectively, this turns 
$\cal T$ into a dimension-7 operator, suppressed only by 3 powers of 
${\bar M}_5$,
enhancing the effect of tree-level graviton exchange.
Despite the IR enhancement in eq.\eq{S(s)},
the graviton contribution still grows with the collider energy.
However, at fixed energy, forward elastic scattering events (i.e.\ small $|t|$)
are statistically as significant as hard scattering events (i.e.\ $t\sim -s$).
This can be contrasted to the $\delta >1$ case,
where the SM background hides graviton effects in forward elastic scattering.

The case $\delta =1$ is rather special, because of
the potential sensitivity to the infrared. 
New infrared-dominated phenomena can occur if, besides gravitons, other
massless (non-derivatively coupled) scalar or vector particles exist in
the bulk. For instance, the tree-level exchange of a massless bulk
scalar particle coupled to a constant source on the Planck brane can give
corrections to parameters on the visible brane proportional to
${\bar M}_5\int dk_5/k_5^2 \sim {\bar M}_5/\mu$, where $\mu$ is the infrared cutoff.
It was argued in ref.~\cite{anto} that this effect can 
destabilize the hierarchy
in the limit of a large compactification radius $R$, whenever $\mu =1/R$.

Suppose that a new U(1) gauge field exist in the bulk. This can mix
in the kinetic term with the hypercharge gauge field, $\epsilon F_{\mu \nu}^Y
 F^{(X)\mu \nu}$, where $\epsilon$ is some effective coupling constant. 
The tree-level exchange of the $X$ gauge KK modes  induces a running
of $\alpha$ linear (and not logarithmic) with energy proportional to 
$\epsilon$. Here we  assume the absence of such scalar or vector
bulk massless particles.

\begin{figure}
$$\includegraphics[width=8cm]{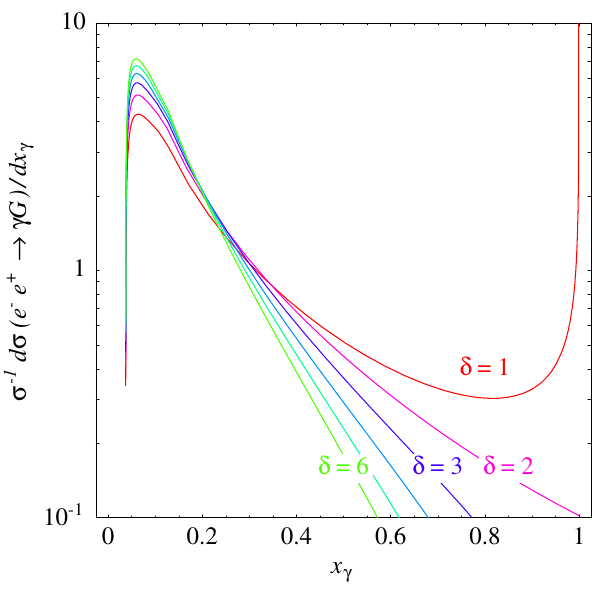}$$
\caption{\label{fig:LEPreal}\em Photon spectrum
($x_\gamma = 2E_\gamma/\sqrt{s}$) in
 $e^+e^-\to \gamma G$ collisions for different 
 numbers of extra dimensions, $\delta=1,2,3,4,5,6$.
 We included the typical cuts performed by LEP collaborations
($p_T^\gamma > 0.0365\sqrt{s}$ and $|\cos\theta_\gamma|<0.95$).}
\end{figure}

\begin{figure}[t]
$$\includegraphics[width=17cm]{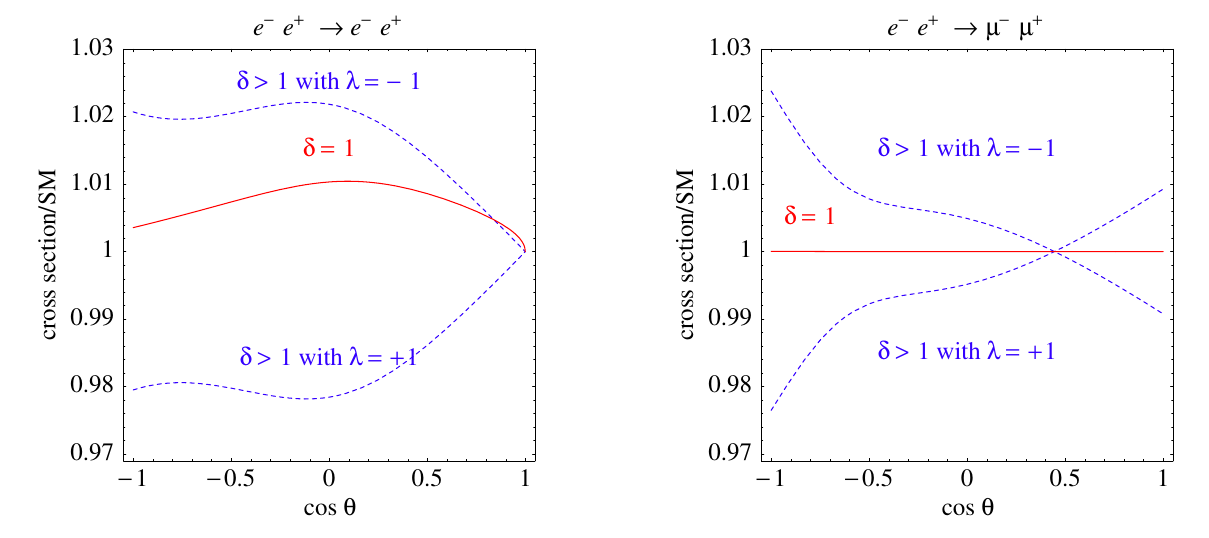}$$
\caption{\label{fig:LEPvirtual}\em Corrections to $e^+e^-\to e^+e^- $ 
(left) and
$e^+e^-\to \mu^+\mu^-$ (right) cross sections
due to tree-level virtual exchange of
gravitons with $\delta = 1$ (red continuous line, for $M_D = 2.5\TeV$)
and $\delta >1$ (blue dashed lines, for ${\cal S} = 8\lambda /(1.25\TeV)^4$
and $\lambda=\pm1$).
In the latter case the result depends on the UV cut-off,
so that not even its sign can be reliably computed: we consider
the two cases of constructive and destructive interference with the SM.
For $\delta=1$ the effect is unambiguously fixed: it
increases the  Bhabha cross section, and gives a negligible correction to inelastic scatterings.
}
\end{figure}

\begin{figure}
$$\includegraphics[width=8cm]{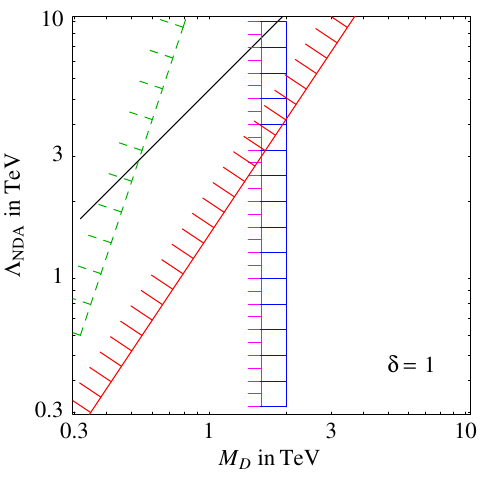}$$
\caption[]{\label{fig:MDL}\em 95\% CL collider bounds on graviton phenomenology in 
the plane $(M_D,\Lambda_{\rm NDA})$ for
$\delta=1$ flat large extra dimension.
The solid black line 
shows the value
of the cut-off $\Lambda_{\rm NDA}$ corresponding to a strongly-interacting
gravitational theory.
The other lines mark the regions excluded by the bounds from 
{\color{blus} graviton emission (vertical blue line)}, 
{\color{purple} tree-level virtual graviton exchange (purple vertical line with short borderlines)}, 
{\color{rossos} graviton loops (red solid line)}, 
{\color{verdes} graviton and gauge boson loops (green dashed 
line)}.}

\end{figure}

\section{Comparison with present data}\label{LEP}
Presently, LEP experiments are the most sensitive probes
of gravitons in one large extra dimension.
We consider three kinds of graviton effects:
1) graviton emission; 2) tree-level virtual exchange of gravitons;
3) one-loop virtual exchange of gravitons.

\paragraph{1) Graviton emission.}
LEP experiments measured the energy and angular spectrum
of $e^+e^- \to \gamma$ + missing energy events at $\sqrt{s}\approx 200\GeV$.
The predicted
$d^2\sigma/dE_\gamma d\cos\theta_\gamma$
(where $E_\gamma$ and $\theta_\gamma$ are the energy and direction of the photon
in the center-of-mass frame) was computed in ref.~\cite{noi} 
and is here reported in appendix A
and plotted in fig.\fig{LEPreal} for different values of $\delta$.
For $\delta >1$ the bounds are dominated by data at small values of $E_\gamma\ll\sqrt{s}$
(where there is some SM background)
because the signal is strongly suppressed at the highest photon energies 
($E_\gamma$ close to $\sqrt{s}/2$),
where the SM background is negligible.
Indeed the differential cross section is divergent when $E_\gamma \to 0$ or
$|\cos \theta_\gamma |\to 1$, because of the familiar
massless electron exchange in the
$t$ channel.
The graviton signal decreases with $E_\gamma$ more slowly 
as the value of $\delta $ is increased.

A novelty specific to the case $\delta=1$ is
the rise around the largest values of $E_\gamma$.
Actually, the differential
cross section
is divergent  for $E_\gamma\to \sqrt{s}/2$ (see fig.\fig{LEPreal}).
Such divergence  corresponds to emission
of KK gravitons with very small mass
and is present because we approximated the dense KK graviton spectrum with a
continuum.  This approximation is valid because
the integrated cross section remains finite.

By imposing $\sigma(e^+e^- \to \gamma G)< 0.01\,\hbox{pb}^{-1}$ at
$\sqrt{s} = 207\GeV$ with the following cuts
$$E_\gamma > 0.7\sqrt{s}, \qquad
p_T^\gamma > 0.0365\sqrt{s},\qquad
|\cos\theta_\gamma|<0.95,$$
we obtain the bound $M_D > 2.4\TeV$.
This  requirement  roughly corresponds  to less than 10 events in all LEP2 experiments.
A precise analysis can only be performed by the experimental collaborations.

\medskip

For $\delta >1$ the process $e^+e^- \to \gamma G$ gives
a more significant probe of graviton emission than $e^+e^-\to ZG$.
This is because the signal is maximal in the kinematical region where
the graviton is heaviest, and the $Z$ mass limits the accessible region
of graviton masses. As previously discussed, the case $\delta=1$ is special,
since a significant contribution to the signal comes from the region of
light gravitons. 
Therefore, for $\delta =1$ one can take full advantage
of the coupling of the $Z$ to electron, which is stronger than the $\gamma$
coupling. 
Appendix A contains our result for the cross section, 
in agreement with previous studies~\cite{ciunciun}.
The L3 limit on the total cross section,
$\sigma(e\bar{e}\to ZG)<0.29\,\hbox{pb}$ at $95\%$ C.L.~\cite{L3}, implies
$M_D >1.4\TeV$ for $\delta=1$.
This limit is subdominant, but using
up-to-date LEP-combined data on the differential
$e\bar{e}\to ZG$ cross section (presently  not available)
might give a probe competitive with other processes.

\paragraph{2) Tree-level virtual graviton exchange.}
As explained in section~\ref{5d}  and illustrated in fig.\fig{LEPvirtual},
virtual effects mediated by tree level exchange
of  gravitons in $\delta =1$ extra dimension are qualitatively different
than in the other cases.
For $\delta >1$ these effects are UV divergent,
so that it is only possible to estimate them by introducing an arbitrary cut-off $\Lambda$,
which becomes more and more important for bigger $\delta$.
Only in the case $\delta = 1$ these effects can be reliably computed,
and used to set a constraint on $M_D$.
In appendix~\ref{cross} we give our results for the cross sections 
of the most interesting processes at electron-positron
and hadron colliders.
When specialized to $\delta >1$ some of our results do not fully agree 
with previous literature.
For $\delta=1$ Bhabha scattering is the only relevant LEP probe~\cite{LEP2}.
We find that it sets the $99\%$ C.L.\ bound $M_D\circa{>}2.4\TeV$.
This value is somewhat higher than the sensitivity of the experiments, 
$M_D=2.0\TeV$ at $99\%$ C.L., because the data show
a moderate preference (almost $2\sigma$)
for an effect with the opposite sign.
A precise analysis can only be done by the experimental collaborations.
(Notice that setting $\delta=1$ in codes developed for studying $\delta >1$ 
is not correct.)

\paragraph{3) One-loop virtual graviton exchange.}
The effects of graviton loops can become more important than tree-level
exchange because they generate effective operators with dimensions lower
than $\cal{T}$. As discussed in ref.~\cite{stru}, especially important
is the operator
\beq\label{eq:Upsilon}
\Lag =c_\Upsilon \Upsilon ,~~~~\Upsilon =\frac{1}{2}
\bigg( \sum_f {\bar f}\gamma_\mu \gamma_5 f\bigg)
\bigg( \sum_f {\bar f}\gamma^\mu \gamma_5 f\bigg) ,
\eeq
where the sum is over quarks and leptons, and the coefficient $c_\Upsilon$
is flavour-universal. Using the rules of na\"{\i}ve-dimensional analysis 
presented in ref.~\cite{stru}, we estimate
\beq
c_\Upsilon =\frac{1}{16\pi^2}\frac{\Lambda_{\rm NDA}^4}{M_5^6},
\label{patuff}
\eeq
where  $\Lambda_{\rm NDA}$ is the UV cutoff.

The different limits on $M_D$ and $\Lambda_{\rm NDA}$ for $\delta =1$
are shown in fig.\fig{MDL}. This figure extends the results of
ref.~\cite{stru} to the case of one extra dimension ($\delta =1$).
Notice that the constraint on $c_\Upsilon$ is the dominant one
for large values of $\Lambda_{\rm NDA}$ and, in particular, it rules out
the case of strongly-interacting gravity at the weak scale. For low
values of $\Lambda_{\rm NDA}$, the dominant constraint comes from
graviton emission and the dimension-8 operator $\cal T$ does not add
new information.

\bigskip

Finally, we comment on graviton contributions to low-energy processes,
which are generally irrelevant for $\delta >1$, but have to be reconsidered
for $\delta =1$\footnote{We thank G.~Isidori for proposing these tests
and for useful discussions.}. First consider the decay $K^+\to\pi^+ G$,
which contributes to the observable 
measured as ${\rm BR}(K^+\to\pi^+{\bar \nu} \nu )=(1.47 ~_{-0.89}^{+1.30})\times 10^{-10}$~\cite{kpi}. 
The na\"{\i}ve estimate, ${\rm BR}(K^+\to\pi^+G )\sim (m_K/M_D)^{2+\delta}$,
suggests that this process is interesting for $\delta=1$.
However a careful analysis gives a different result.
The part of the chiral Lagrangian bilinear in the charged $K$ and $\pi$ mesons is
\beq
\Lag =a_{ij}(D_\mu \phi_i)^\dagger (D^\mu \phi_j )-m^2_i \phi^\dagger_i
\phi_i .
\eeq
Here $\phi =(\pi^-,K^-)$, $m^2=(m^2_\pi,m^2_K)$, $a_{11}=a_{22}=1$,
$a_{12}=a_{21}=-\sqrt{2}G_Ff_\pi^2g_8\sin\theta_{\rm C} $, and $|g_8|=5.1$
is the octet coupling. With a field redefinition, we can simultaneously
diagonalize both the kinetic and mass terms. The graviton
is coupled to the current $T_{\mu \nu}=g_{\mu \nu}\Lag-(\partial \Lag/
\partial\partial^\mu \phi_i )\partial_\nu \phi_i$, which is also diagonal.
Therefore (in contrast with axion emission), one finds no tree-level
contribution to $K^+\to\pi^+ G$~\cite{bijnens}. 
Loop effects exist
but, given the LEP limits previously obtained, they are very small, since
they are suppressed by a factor $(m_K/M_5)^3/(4\pi)^4 \sim
(2~{\rm TeV}/M_5)^3~6\times 10^{-16}$. The processes $(q{\bar q})\to G\gamma$
are possible at tree level, but no interesting bounds can be extracted
from $J/\psi$ or $\Upsilon$ decays~\cite{bijnens}.

\section{Graviton signals at the LHC}\label{LHC}

Collider signatures of KK gravitons occur through real-graviton
emission and virtual-graviton effects. 
Real-emission signals give clean signatures with
jets
or leptons plus missing energy, a class of signals very carefully
studied in the framework of supersymmetry. 
A hard cut
on missing energy removes QCD backgrounds very efficiently and, even
for jet final states, leaves us mostly with $W$ and $Z$ boson
production processes in the Standard Model. These background processes are fairly well
understood theoretically and have small rates. In this
section we will also explore virtual-graviton search strategies in
detail. These signals come from tree-level virtual-graviton
exchange (the dimension-8 $\cal T$ operator)
and one-loop graviton contributions, which lead to the dimension-6
axial-axial
operator $\Upsilon$.
Because virtual-graviton effects have no golden cut to
efficiently suppress Standard Model backgrounds, our focus moves from
the computation of the signal rates to the estimate of the error on
the background processes.\medskip

We move beyond a one-parameter description in which one
identifies the $D$ dimensional Planck scale $M_D$ with the cutoff
$\Lambda$ of the theory. We apply a cutoff $\Lambda$ to the graviton-emission
and virtual-graviton exchange rates,
setting the partonic cross section to zero when
$\hat{s}>\Lambda^2$.\footnote{Ref.~\cite{hinchliffe} suppressed the
cross sections in a slightly different way, 
by adding in a factor $\Lambda^4/\hat{s}^2$ for
$\hat{s}>\Lambda^2$. Our
results and those in~\cite{hinchliffe} are in good agreement.}
 This cutoff is important for the LHC
because with a hadronic collider energy of $14\tev$ we expect to probe
a wide range of energies. 
Separating the cutoff from the Planck mass allows us to investigate the
behavior of the effective KK theory for LHC observables.
Unless explicitly stated otherwise, we 
combine the statistical significance $S/\sqrt{B}$ and the
systematical significance $S/\Delta B$ in quadrature and we require a
 $5\sigma$ effect. 

The same cutoff $\Lambda$ is used to regulate the ultraviolet divergences
that appears in operators induced by virtual-graviton exchange.
These divergences are expected to be removed by the
ultraviolet completion of our effective KK theory. Because we do not
know this ultraviolet completion, the dependence on the cutoff
$\Lambda$ parametrizes our ignorance of ultraviolet effects. 
\medskip

Effects from virtual graviton exchange are described in terms of $M_D$, 
$\Lambda$ and $\delta$. 
The dependence is defined by the coefficients $c_{\cal T}$
and $c_\Upsilon$ of the dimension-8 and dimension-6 operators, obtained
in modified na\"{\i}ve dimensional analysis (NDA)~\cite{stru}
\beq
c_{\cal T} =\left\{\begin{array}{ll }\displaystyle
 \frac{\pi^{\delta /2}}{\Gamma (\delta /2)} \;
               \frac{\Lambda_{\rm NDA}^{\delta -2}}{M_D^{\delta +2}}      &  {\rm for}~\delta >2  \\
\displaystyle
\frac{\pi}{M_D^4}\ln \frac{\Lambda_{\rm NDA}^2}{E^2}      &   {\rm for}~\delta =2
\end{array}\right.,\qquad
c_\Upsilon = \frac{1}{16} \;
               \frac{\pi^{\delta -2}}{\Gamma^2 (\delta /2)} \;
               \left( \frac{\Lambda_{\rm NDA}^{\delta +1}}{M_D^{\delta +2}}
               \right)^2
\label{eq:op_lhc}
\eeq
where $E$ is a typical partonic energy of the process. The expressions for
$\delta =1$ are given by eqs.~(\ref{eq:S(s)}) and~(\ref{patuff}). 

The ratio $\Lambda/M_D$ controls how strongly gravity is coupled,
i.e.\ the relative importance of tree and loop effects. Even in string models of quantum gravity
this remains an unknown parameter, controlled by the vev of the dilaton.
The maximal value of $\Lambda/M_D$ corresponds to the case of
strongly-coupled gravity, where all loops give comparable contributions.
Na\"{\i}ve dimensional analysis estimates it to be~\cite{stru}
\beq
\Lambda \circa{<}\Lambda_S \equiv \left[ 16 \, \pi^{\frac{4-\delta}{2}} \, \Gamma
\left( \frac{\delta}{2}\right) \right]^\frac{1}{2+\delta}M_D .
\label{eq:lambda_s}
\eeq
For real graviton emission, the dependence on the ultraviolet cutoff
$\Lambda$ measures the fraction of the signal coming from different 
regions of the KK mass spectrum, while the matrix element itself is
independent of $\Lambda$.  Again, we limit the parton energies to
$\sqrt{\hat{s}} < \Lambda \circa{<} \Lambda_S$. 

We stress that care should be applied when different signals are compared,
since the cutoff $\Lambda$ will be in general different for different
processes: order one factors can be fixed only after understanding quantum gravity.


\subsection{Graviton emission}

Two well-suited search channels
for real-graviton emission 
are a single hard jet plus missing
energy~\cite{noi} and Drell--Yan lepton pairs plus missing
energy~\cite{dave}. The latter is a particularly clean channel, but
its reach has shown to be somewhat less than the single jet
counterpart's. We therefore concentrate on the process $pp \to {\rm
G+ jet}$.\medskip

We perform a parton-level analysis, including the dominant irreducible
background $pp \to Z+{\rm jet}$ where the $Z$ decays to
neutrinos. Apart from a basic acceptance cut for the jet, $|\eta_j|<4.5$,
we apply a variable transverse momentum cut: $p_{T,j}>\Lambda/4$ for
$\Lambda<4\tev$ and $p_{T,j}>1\tev$ otherwise. To first approximation the
signal and the dominant background processes will lead to a
back-to-back signature of the jet with the invisible KK graviton or
$Z$ boson, respectively. Therefore, there is no need to apply a hard
missing-energy cut beyond the triggering requirements --- for each
event we expect $\Emisst \sim p_{T,j}$. A minimum $\Emisst$ cut could be
added, but would only increase the uncertainty because of the $\Emisst$
smearing of the detector and the appropriate calibration. The second
largest background arises from $pp \to W+{\rm jet}$, where we lose the
lepton from the $W$ decay because it is too soft or too far
forward. Even though the rate before acceptance cuts is much larger
than the $Z+{\rm jet}$ rate, after cuts it only leads to a
$\sim 10\%$ correction to the $Z+{\rm jet}$ background.\medskip

\begin{figure}[t]
\centering
\includegraphics[width=9cm]{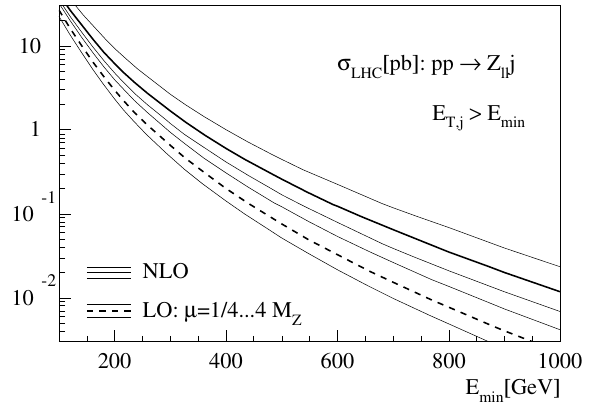}
\vspace{0mm}
\caption[]{\em The leading order and NLO cross sections for the process $pp
\to Z+{\rm jet}$ with a subsequent $Z$
decay into electrons or muons~\cite{Giele:dj}. 
The factorization and renormalization scales
are set equal and varied around a common central value. The only cut
applied is the shown minimum cut on the transverse jet momentum.}
\label{fig:lhc_nlo}
\end{figure}

The nature of the graviton signal does not permit any kind of
simple side-bin analysis. All information is encoded in the hard
$p_{T,j}$ spectrum of the signal plus the $Z+{\rm jet}$
background. The NLO correction to the backgrounds $pp \to Z/W+{\rm
jet}$ are known~\cite{Giele:dj}. We show the leading order as well as
the NLO cross sections as a function of the $p_{T,j}^{\rm min}$ cut in
fig.~\ref{fig:lhc_nlo}.
If we do not apply any cut,
we obtain a $K$ factor of $1.6$ and a remaining NLO scale variation of
the order of $25 \%$. For a large $p_{T,j}$ cut the relative
correction increases to $K \sim 2$ with a remaining NLO scale
uncertainty of almost $100 \%$. This does not even take into account
the fact that we have two scales present in the process, namely the
weak boson mass $m_Z$ and the transverse momentum of the jet --- in
principle we would have to resum large logarithms $\log
p_{T,j}/M_Z$. Therefore we can immediately conclude that trying to
extract the KK signal out of the $Z+{\rm jet}$ background using the
theoretical prediction of the $p_{T,j}$ spectrum of the $Z+{\rm jet}$
final state will have to cope with very large systematic errors and it is
therefore pointless.

Instead, we normalize the $Z+{\rm jet}$ rate using $Z$ decays to
electrons and muons, as suggested in ref.~\cite{noi}. 
Since this final state can be fully
reconstructed, we can even mimic signal cuts and look for effects from
large logarithms induced by the $p_{T,j}$ cut in the actual
analysis. The remaining statistical uncertainty on the background
measurement can be estimated from fig.~\ref{fig:lhc_nlo}. Using only
$10 \fb^{-1}$ of data and a $p_{T,j}$ cut of $600 \gev$, we are left
with $S=1000$ events for one lepton flavor, which gives us a
statistical uncertainty of $1/\sqrt{S} \sim 3\%$ per lepton
flavor. Combining both the electron and the muon final state it seems
to be conservative to assume a systematic uncertainty of $5 \%$ on the
prediction of the $Z+{\rm jet}$ background cross section including
detector effects.

The $W+{\rm jet}$ background cannot be reconstructed, because the
three-quark final state with a hadronic $W$ decay will be dominated by
the large QCD three-jet continuum. However, 
the NLO corrections scale with the $Z+{\rm jet}$ production
process; \ie the $W+{\rm jet}$ rate can be estimated from the measured
$Z+{\rm jet}$ rate. 
Since the $W+{\rm jet}$ rate gives a contribution to
the total background of at most $10\%$, and is affected by
an uncertainty of less than $20\%$, 
we can essentially neglect this contribution, which is taken into account 
in 
the $5\%$ quoted uncertainty.\medskip

\begin{figure}[t]
\centering
\includegraphics[width=16cm]{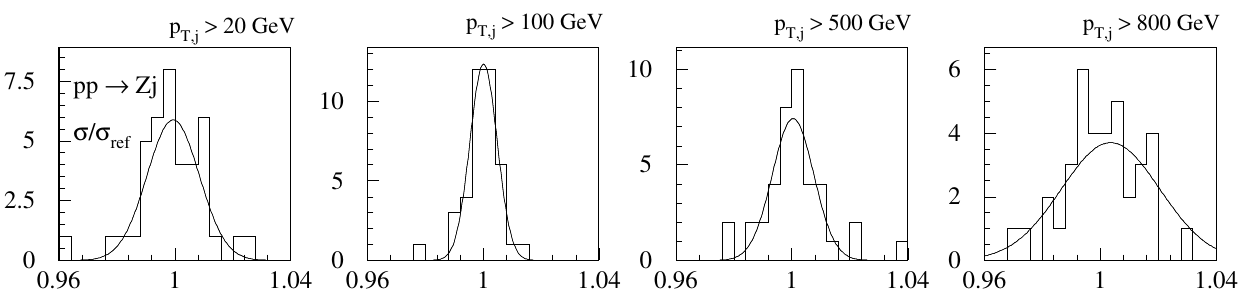}
\vspace{0mm}
\caption[]{\em The range of $Z+{\rm jet}$ background cross
sections, due to the uncertainty in the parton densities. The histograms 
show the deviation $\sigma/\sigma_{\rm Cteq6.1M}$.
The set of 40 parton densities included in the histograms is described in
ref.~\cite{Stump:2003yu}.}
\label{fig:lhc_pdf}
\end{figure}

For completeness, we study the uncertainty caused by the parton
densities. This source of uncertainty can be possibly reduced through a
complete two-loop analysis with HERA and hadron collider jet data, but
it should not just be neglected. For different $p_{T,j}$ cuts we show
histograms using a consistent set of different Cteq6.1 parton
densities~\cite{Stump:2003yu} in fig.~\ref{fig:lhc_pdf}. 
As expected, we observe a very
narrow distribution if we require $p_{T,j}>100 \gev$, pushing the
parton momentum fraction into the well determined region $x \sim
1/100$. In the region of larger transverse jet momenta, where we can
observe the graviton radiation signal, the probed $x$ values become larger,
but the systematic uncertainty on the cross section prediction
inherited from uncertainties in the parton density input does not grow
beyond $2\%$, the maximum width we observe in fig.~\ref{fig:lhc_pdf}.
This small uncertainty can be traced back to the parton
kinematics: while the uncertainty on the gluon densities for $x>0.5$
is indeed very large, it stays small for valence quarks~\cite{Stump:2003yu}.
We include at maximum one gluon in the initial state. When the
gluon luminosity drops sharply for large $x$ values, the mixed $qg$
initial state will be pushed to asymmetric $x_g \ll x_q$
configurations, which do not suffer from large uncertainties. Last but
not least, this theoretical uncertainty is of no concern if we measure
the $Z+{\rm jet}$ background, but even if we were to combine the
measured background rate and its theoretical prediction the parton
densities would be no reason to worry.\bigskip

\begin{figure}[t]
\centering
\includegraphics[width=9cm]{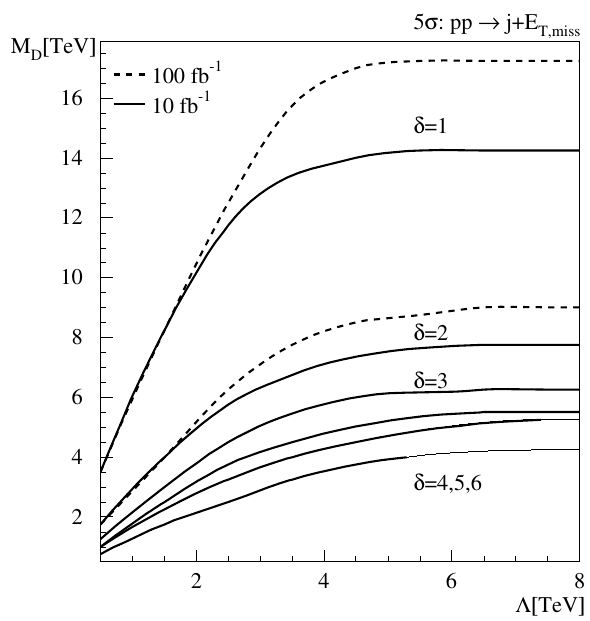}
\vspace{0mm}
\caption[]{\em The LHC reach in the search for graviton emission. The
number of extra dimensions $\delta$ is varied from 1 to 6, in the
lines from top to bottom. The switch from thick to thin lines indicates
the end of the region $\Lambda<\Lambda_S$.}
\label{fig:lhc_emission}
\end{figure}

In fig.~\ref{fig:lhc_emission} we show the discovery reach for
graviton plus jet production at the LHC, assuming an integrated
luminosity of $10 \fb^{-1}$ and $100 \fb^{-1}$. The area to the right
and below the lines will lead to a $5 \sigma$ signal with combined
systematical and statistical errors. Large values of $M_D$ lead to a
direct suppression of the graviton coupling and thereby a suppression
of the signal rate. Smaller cutoffs remove a larger fraction of the
graviton mass spectrum we integrate over.\smallskip

The plateau reached by the curves at large $\Lambda$ indicates the region
where the signal becomes independent of the ultraviolet cutoff. If the
onset of the plateau starts when $\Lambda$ is in the perturbative
regime ($\Lambda <\Lambda_S$), the use of the effective theory, obtained
by linearizing Einstein gravity, is well justified (for the
values of $M_D$ probed). In this case, the
contribution to the signal coming from very energetic gravitons is strongly
suppressed by parton luminosity. On the other hand, 
if the plateau is reached only for $\Lambda >\Lambda_S$, the signal is 
sensitive
to the unknown physics from the ultraviolet completion of the
effective theory. This happens for $\delta \gtrsim
5$, as shown in fig.~\ref{fig:lhc_emission}. 
Over some fraction of the
discovery contour shown in fig.~\ref{fig:lhc_emission} we find $M_D
\sim \Lambda \sim \sqrt{\hat{s}}$. 
While the one-graviton production cross section remains finite
even for large cutoff values, the complete $\Emisst$ signal will
consist of an uncalculable increasing fraction of multi-graviton emission. 

For a smaller number of extra dimensions, the graviton mass
spectrum is dominated by lighter modes, since the graviton mass
distribution is proportional to $m^{\delta -1}dm$ (and, in particular,
becomes flat for $\delta =1$).  This means that, for small $\delta$,
the cutoff will have a less dramatic effect: the transition to the
plateau is more abrupt and the discovery reach becomes independent of
$\Lambda$ already for $\Lambda \ll \Lambda_S$. Moreover, since the
cross section is proportional to $M_D^{-2-\delta}$, we can
probe much larger values of $M_D$ for a small number of extra
dimensions.  
For larger values of $\delta$, the signal is dominated by gravitons
with the largest possible energy, and therefore the reach on $M_D$
critically depends on $\Lambda$, as shown in fig.~\ref{fig:lhc_emission}.

We recall that the curves
in fig.~\ref{fig:lhc_emission} show the upper limit on $M_D$ that can be 
reached by LHC.
The successful theoretical description of the LHC reach
is based on the separation of the two scales $\sqrt{\hat{s}} <
M_D$, which works very well for the maximum $M_D$ values which can 
be probed. A lower bound on the range of $M_D$ which can be probed is 
given by the theoretical
requirement $\Lambda <\Lambda_S$ and the condition that $\Lambda$ has to 
be larger than the typical partonic collider energy, thereby allowing
for the observed saturation behavior. For smaller values of $M_D$
the requirement that gravity cannot
become strongly interacting can lead a theoretical uncertainty
$d\sigma(M_D,\delta;\Lambda)/d\Lambda \ne 0$ that has to be properly taken into
account.
Looking at the different values of $\delta$ we conclude that LHC
searches for KK emission become more probing and less
theoretically uncertain the smaller the number of extra
dimensions.\medskip

The improvement of the LHC reach with a luminosity increase is linked
to either a dominating systematical or statistical error: for small
cutoff values $\Lambda$, the transverse momentum cut is small as well,
leaving us with a large background rate. The systematical error on
this large background rate determines the reach, so a luminosity
increase will only slightly improve the reach. Taking into account detector
effects, a purely high-luminosity sample is likely to be inferior to
its low-luminosity counterpart. Once we probe larger values of
$\Lambda$, which is possible for a small number of extra dimensions,
the background suppression cut tightens and a luminosity increase does
have a sizeable effect, as we can see in the $\delta=1,2$ cases in
fig.~\ref{fig:lhc_emission}.\smallskip

We have confronted the real graviton emission with the Standard
Model process $pp \to Z+{\rm jet}$, but new backgrounds can arise from
the unknown dynamics at the cutoff scale. Although tree-level graviton
exchange does not directly contribute to the dominant parton
subprocess $qg \to qZ$, in general there could be other effective
operators involving neutrinos or $Z$ bosons, which can affect our
background measurement. While we know from LEP that on the $Z$ pole
virtual graviton exchange cannot have a large effect, unknown neutrino
operators could mimic a KK mediated deviation from the Standard Model
$pp \to Z+{\rm jet}$ prediction.

\subsection{Tree-level virtual graviton exchange}

There are three most promising 
candidate channels to study tree-level virtual graviton effects:
Drell--Yan lepton pairs, photon pairs and two jets. 
The dimension-8 operator $\cal{T}$ induced by tree-level graviton
exchange can interfere with Standard
Model contributions for $\delta >1$ and with the elastic contribution
for $\delta =1$.
Therefore, it can lead to effects suppressed by only
one power of ${\cal S} \sim M_D^{-2-\delta}$. Because we are dealing with
interfering amplitudes, there
will be no golden cut to distinguish signal and backgrounds; at most
we expect significant changes in some distribution. Instead of the
transverse momentum distribution which we have utilized for the real emission, we now study
the invariant mass of the final state leptons or photons or jets to
measure a deviation from the Standard Model behavior. In contrast to the
real emission case, there is no independent background measurement, but
the Drell--Yan $\mll$ spectrum can at least be normalized at the low
end just above the $Z$ pole, where the heavy KK states
should not contribute. 

In earlier works the reaches of the two lepton and the two photon
channels have been found to be comparable~\cite{Kingman,atlas_dy}. The
diphoton processes, however, suffers from several disadvantages: the
total rate is only known at NLO~\cite{binoth}, with an estimated error of about 
$20$--$30\%$. A normalization of the invariant mass distribution in
the small $M_{\gamma \gamma}$ region might suffer from soft-photon
radiation on the one hand, and its extrapolation is affected by 
large logarithms
on the other hand. Finally the error due to QCD uncertainties of
misidentified photons will be at least of the order of $10\%$ to
$20\%$ for the extracted photon rate, well above the estimated error
for the Drell--Yan process~\cite{binoth}.\medskip

\begin{figure}[t]
\centering
\includegraphics[width=7.2cm]{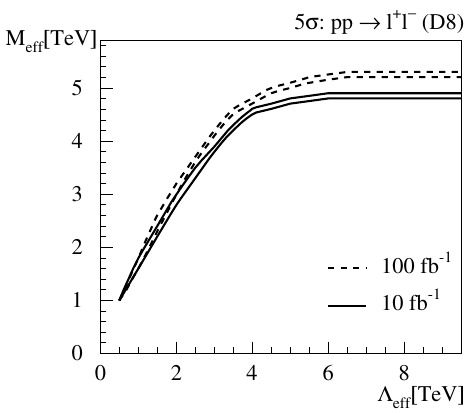}\hspace*{5mm}
\includegraphics[width=9cm]{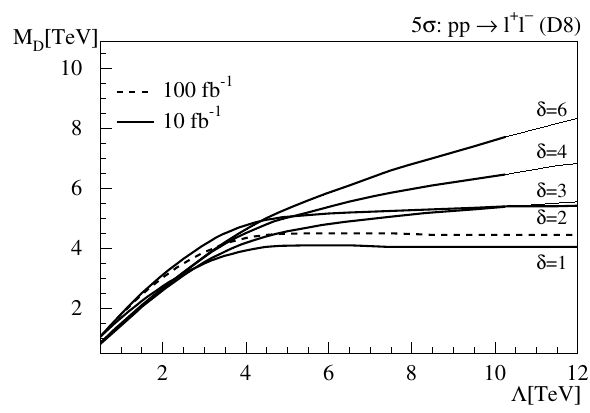} 
\vspace{-3mm}
\caption[]{\em The LHC reach in the search for virtual graviton
effects through the dimension-8 operator $\cal T$ in Drell-Yan lepton
pair production. Left: the result in terms
of the coefficient $c_{\cal T}$ of the dimension-8 operator, with 
$c_{\cal T} = \pm 8/\meff^4$ (the sign does not affect the result). The
reach is shown for the $q\bar{q}$ initial state only (lower $\meff$)
and including the $gg$ initial state (higher $\meff$). The cutoff
$\Lambda_{\rm eff}$ acts on the partonic collider energy and defines
the minimum $\mll$ cut exactly like $\Lambda$ in the extra-dimensions
parametrization. Right: the same result in terms of the extra-dimension
parameters $\delta$, $\Lambda$, and $M_D$.
The switch from thick to thin lines indicates the
end of the region $\Lambda<\Lambda_S(M_D)$.}
\label{fig:lhc_d8}
\end{figure}

Again, we only require basic cuts for the leptons $|\eta_\ell|<2.5$
and $p_{T,\ell}>50 \gev$, plus a variable minimum cut $\mll>\Lambda/2$
for $\Lambda<4 \tev$ and $\mll>2 \tev$ otherwise. The $5 \sigma$ reach of
this channel is illustrated in fig.~\ref{fig:lhc_d8}.
We start by showing our results, in the left panel of the figure, 
by using the model-independent
parameters $\meff$ (where $c_{\cal T} \equiv 8/\meff^4$) and $\Lambda$, the
ultraviolet cutoff. In this case, $\Lambda$ corresponds to the maximum
allowed value of the two-lepton invariant mass $\mll$. 

In contrast to the
background, the KK signal involves two parton subprocesses: the usual
Drell--Yan type diagram with an $s$ channel KK exchange, and a gluon initiated
process. The latter will have a large effect on the total rate
at the LHC, but the $\mll$ cut selects the large parton $x$ regime
where initial state quarks dominate (see left panel of fig.~\ref{fig:lhc_d8}).
The interference between KK and Standard Model
contributions to the Drell-Yan channel
is strongly suppressed by a combination of couplings and phase space
factors when we integrate over the entire phase space. Therefore,
the rate for the KK effects is not proportional to $c_{\cal T}$,
but to $c_{\cal T}^2$. Indeed the curves for positive and negative values
of $c_{\cal T}$ in the left panel of fig.~\ref{fig:lhc_d8} are nearly
identical. Taking into account only the interference term would reduce
the LHC reach to $\meff <2$~TeV.
We
note, however, that it might still be possible to enhance the
reach applying a more sophisticated set of cuts to increase the
interference effects.
As long as $\Lambda$ is larger than about 4~TeV, the reach shown in
the left panel of fig.~\ref{fig:lhc_d8} is independent of $\Lambda$,
since the condition $\mll <\Lambda$ does not reduce the signal.
\medskip

Next, we show in the right panel of
fig.~\ref{fig:lhc_d8} the same result expressing $c_{\cal T}$
in terms of $\Lambda$ and $M_D$, as given in eq.~(\ref{eq:op_lhc}).
For $\delta >2$, the LHC reach on $M_D$ grows with $\Lambda$, as
an effect of this particular parametrization. 
The form of ${\cal S}$ will lead to a scaling $M_D^{\rm max} \propto
\Lambda^{(\delta-2)/(\delta+2)}$ for $\Lambda \gtrsim 1\tev$. 
The growth is limited by the requirement $\Lambda <\Lambda_S$ which 
implies $\Lambda < (\sqrt{2} \pi)^{1/2} \meff^{\rm max}\simeq 10$~TeV,
for any $\delta >2$, where $\meff^{\rm max}$ is the reach on $\meff$,
shown in fig.~\ref{fig:lhc_d8}.
Because the interference between KK and Standard
Model contributions is suppressed, the numerical impact of a shift
$\Lambda_S=M_D \to 4 M_D$ on the total cross section could be as large
as a factor $4^{2(\delta-2)}$. For $\delta=6$ we find a change of the
cross section by more than a factor $10^5$ for fixed $M_D=4\tev$ and a
fixed kinematic cut $\mll>1\tev$. The total cross section, however, is
extremely sensitive to changes in $\Lambda$ and also in $M_D$. If we
translate the factor $4^{2(\delta-2)}$ into a compensating shift in
$M_D$ we only get a shift by a factor $(4)^{(\delta-2)/(\delta+2)}$,
which for $\delta=6$ means $M_D \to 2 M_D$. 
\medskip

For $\delta=2$ the
cutoff dependence of $c_{\cal T}$ 
is only logarithmic, and $c_{\cal T}$ is finite
for $\delta =1$. 
In fig.~\ref{fig:lhc_d8} we see that, for these two cases, the LHC reach is
practically independent of $\Lambda$, as soon as $\Lambda >4$~TeV, when
the condition $\mll <\Lambda$ no longer affects the signal.
Therefore, for small $\delta$ (and especially for $\delta =1$), the
coefficient is less sensitive to the ultraviolet behavior of the
theory, here parametrized by $\Lambda$. 
This means that 
the search for tree-level virtual-graviton effects,
although
theoretically more under control, covers a smaller portion of the
$M_D$--$\Lambda$ region, as the value of $\delta$ is decreased. 
This is in contrast to real-graviton emission, where the reach is enhanced for
small $\delta$.

\begin{figure}[t]
\centering
\includegraphics[width=9cm]{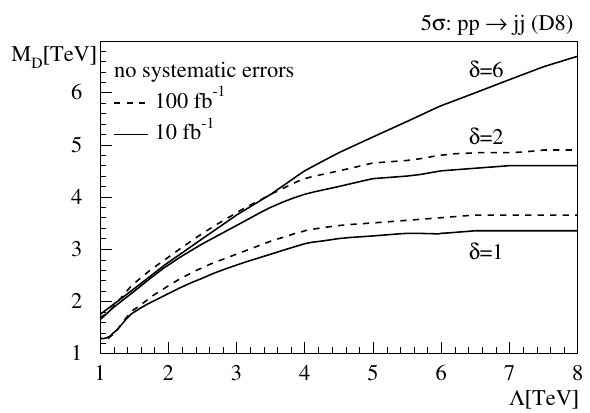}
\vspace{0mm}
\caption[]{\em The LHC reach in the search for tree-level virtual-graviton effects 
through
the dimension-8 operator ${\cal T}$ in 
two-jet process. For illustrative purposes, the systematic errors are
set to zero, even though they will dominate and even nullify the
reach.}
\label{fig:lhc_jj}
\end{figure}

The two-jet channel is of course tempting at the LHC, just because of
its large rate. However, this rate also comes with a large uncertainty,
not only on the total rate but also on distributions. The invariant
mass spectrum for example is a likely candidate to measure parton
distribution functions, which means that even if we knew its
normalization (which we do not know) it would be hard to extract any
information from it. In fig.~\ref{fig:lhc_jj} we show an example for
discovery reach of the LHC looking for dimension-8 virtual graviton effects for
two jet production {\sl in the limit of zero systematical
uncertainty}. We see that the reach is similar to the corresponding Drell--Yan
process; however, we find $S/B$ well below $10^{-2}$, which means that
we do not even have to estimate a systematical error on the QCD two
jet production --- any reasonable number will return a discovery potential
worse than in the lepton pair channel.

\subsection{One-loop virtual graviton exchange}

\begin{figure}[t]
\centering
\includegraphics[width=7.2cm]{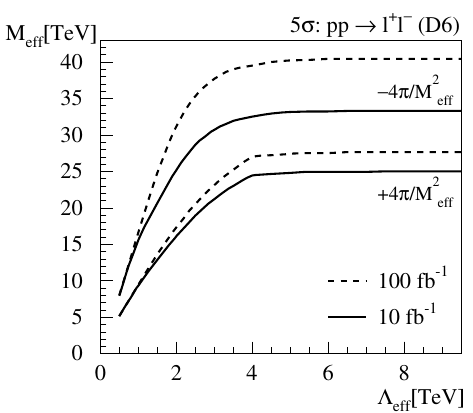} \hspace*{5mm}
\includegraphics[width=9cm]{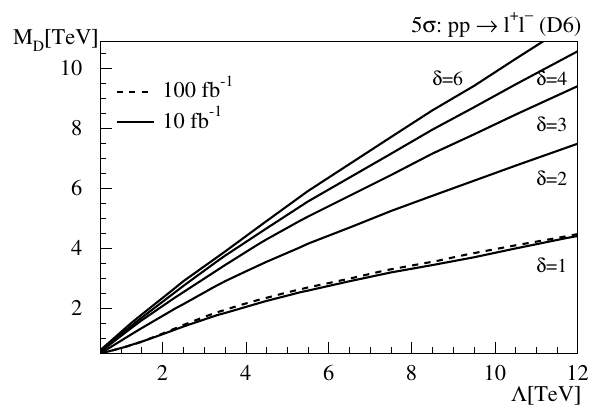} 
\vspace{-8mm}
\caption[]{\em The LHC reach in the search for graviton-loop effects
through the dimension-6 operator $\Upsilon$ in Drell-Yan lepton-pair
production. The cutoff $\Lambda_{\rm eff}$ acts on the partonic
collider energy and defines the minimum ${\mll}$ cut exactly like
$\Lambda$ in the extra-dimension parametrization. Left: the results
in terms of the parametrization $c_\Upsilon = \pm
4\pi/\meff^2$. Right: the reach in terms of the parametrization in
eq.~(\ref{eq:op_lhc}), for a positive value of $c_\Upsilon$.}
\label{fig:lhc_d6}
\end{figure}

Graviton loops induce a dimension-6 operator $\Upsilon$~\cite{stru,oleg}.  
We perform the same analysis
of the Drell--Yan lepton pairs at the
LHC as for the dimension-8 operator $\cal T$.
First, we determine the LHC reach in terms of the 
parametrization $c_\Upsilon = \pm 4\pi/\meff^2$. Because of the
cancellation between interference effects and the dimension-6 operator
squared, a positive sign $c_\Upsilon$  has a reduced effect on the
Drell--Yan cross section. From fig.~\ref{fig:lhc_d6} we observe that 
the cutoff
dependence disappears for 
$\Lambda >
4\TeV$, which is only slightly above the typical $\mll$ values at the LHC
and well below the maximum values $\meff \sim 25$--$40\TeV$. 
The two scales are well separated, and the result is fairly
independent of the
cutoff.
We can translate the maximum range in terms of the
dimension-8 operator $c_{\cal T}^2=(8/\meff^4)^2$ into the range for
the dimension-6 operator $c_\Upsilon = 4\pi/\meff^2$, assuming the
typical partonic collider energy is $\mll \sim 2\TeV$.\bigskip

As a second step we interpret the same result in terms of the
extra-dimensional
effective theory, using the definition of the
dimension-6 operator coefficient $c_\Upsilon$ in eq.~(\ref{eq:op_lhc}). 
The dependence on $\Lambda$ of $c_\Upsilon \sim 
\Lambda^{2\delta+2}/M_D^{2\delta+4}$ is steeper than for the real 
emission and the dimension-8
cases.
Moreover, comparing the two parametrizations of the dimension-6 operator
for $M_D = \Lambda = \meff$ we find a considerably smaller
prefactor for $c_\Upsilon$ for the extra dimension
ansatz. 
Correspondingly, the $5 \sigma$ discovery curves in
fig.~\ref{fig:lhc_d6}, shown for positive $c_\Upsilon$, 
only barely cross the diagonal $\Lambda<M_D$.
In contrast to the dimension-8 operator the dimension-6 analysis will
have basically no reach at the LHC if we limit ourselves to
$\Lambda<M_D$, but surpasses the dimension-8 operator for $M_D <
\Lambda < \Lambda_S$.
This is consistent with the notion that loop effects become overwhelming
when $\Lambda$ is large, and the effective gravitational coupling 
becomes sufficiently strong.

The reach on $M_D$ shown in fig.~\ref{fig:lhc_d6} grows with 
$\Lambda$ almost linearly, $M_D^{\rm max} \propto
\Lambda^{(\delta+1)/(\delta+2)}$, until it is overcome by the condition
$\Lambda <\Lambda_S$. This happens for $\Lambda =2\pi^{1/2} \meff^{\rm max}
\sim 89\TeV$, for any $\delta$. This means that searches for the
operator $\Upsilon$ can very efficiently cover the case of strong gravity
($\Lambda \gtrsim M_D$) in a large range of $M_D$ values, namely up
to $M_D^{\rm max}=(16,33,47,56,62,66)\TeV$ for $\delta =(1,\dots ,6)$.
This typical scale of $M_D^{\rm max}$ is set by the
maximum value of $\meff$ which we can probe at the LHC. 
This corresponds to the hierarchy of scales in
the dimension-8 case, where the typical scale is set by $\meff \sim
5\TeV$, the reach in $M_D$ is somewhat larger (depending on $\delta$),
and $\Lambda$ is fixed by the condition $\Lambda<\Lambda_S(M_D)$. The
collider energy as a scale has no impact once $s/\meff^2$ fixes the
reach in $\meff$.

\section{Conclusions}

We have reconsidered the physics of KK graviton excitations 
in theories with
$\delta$ large extra dimensions. The high-energy
and low-energy ends of the KK spectrum lead to different effects.
\begin{itemize}
\item The heaviest KK modes affect high-energy experiments
in a way that does not depend on the detailed shape of the
compactification. However, while graviton emission is a truly
model-independent test, virtual-graviton exchange for $\delta \ge 2$
cannot be fully computed without knowledge of the UV completion.
\item The lightest KK modes affect Newtonian gravity at measurable distances
and astrophysical processes
in a fully computable way
that however depends on the detailed shape of the extra dimensions.
In models with $\delta$ flat extra dimensions with equal radii, the KK masses
are $m\sim M_D (M_D/M_{\rm Pl})^{1/\delta}$, 
leading to conflict with observations for $\delta \le 3$ and $M_D<3$~TeV.

\end{itemize}
Many previous works studied high-energy signals of $\delta=2,3$ extra 
dimensions, tacitly
assuming the existence of 
some appropriate non-minimal shape that 
removes the light KK modes.
The case $\delta=1$ was ignored, 
maybe because it is less trivial to imagine an appropriate non-minimal shape 
for a one-dimensional extra dimension, 
or maybe just because of psychological reasons:
in the flat limit the case $\delta=1$ contradicts even ordinary experience.

\smallskip

In section~\ref{seccaz} we discussed how making the long extra dimension slightly warped
introduces an extra mass $\mu$ to the spectrum of KK excitations
that allows to make all states heavier than about $50\MeV$,
avoiding problems with observations.
Warping also makes each KK more strongly coupled:
the combined effect of these two factors leaves unaffected high-energy experiments 
which cannot resolve the individual KK modes.

The explicit model motivates our later study of high-energy signals,
which 
do not depend on the specific mechanism introduced to cure low-energy problems.
In section~\ref{5d} we studied collider signals of a
single gravitational large extra dimension, findings
a few peculiarities.
\begin{itemize}
\item[1)]  Unlike $\delta\ge 2$,
for $\delta=1$ the graviton-emission cross sections are not dominated by 
events with the largest kinematically allowed graviton energy.
LEP measurements of  $e^+e^-\to \gamma G$ events with large $E_\gamma\circa{>}M_Z$
give the constraint $M_D \circa{>} 2.4\TeV$ on the 5-dimensional Planck mass
(normalized as in ref.~\cite{noi}).
Furthermore, unpublished $e^+e^- \to Z G$ data might give a competitive test.

\item[2)]
For $\delta \ge 2$ tree-level virtual graviton exchange
gives the dimension-8 operator ${\cal T}$ of eq.\eq{T}
which has an
uncomputable UV-divergent coefficient~\cite{noi}.
For $\delta=1$ tree-level virtual graviton exchange
gives a dimension-7 fully computable amplitude,
which interferes with the SM amplitude only in elastic channels.
The dominant effect at LEP is an increase of the $e^+e^-\to e^+e^-$ cross section,
with a peculiar scattering-angle dependence.
The present constraint  is $M_D\circa{>}2.4\TeV$.

\item[3)] One loop graviton exchange gives the dimension-6 operator $\Upsilon$ of eq.\eq{Upsilon};
its coefficient is UV-divergent coefficient for any $\delta$, including $\delta=1$.

\end{itemize}

\medskip

Finally we reconsidered how the sensitivity of LHC searches
to gravitons in $\delta$ large extra dimensions
depends on the cut-off $\Lambda$ of the theory.
Previous analyses assumed $\Lambda = M_D$, which
fixes in an arbitrary way the relative merit of the different signals discussed above.
In appendix~A we give the cross sections 
of graviton-induced hadronic processes relevant for LHC studies
(with finding that do not fully agree with previous 
results valid for $\delta\ge 2$).

Searches for real-graviton emission, tree-level graviton exchange, and
graviton-loop effects give complementary information. For graviton emission,
we have focused on the jet and missing energy channel. The case $\delta =1$
is particularly interesting, since LHC can probe values of $M_D$ up to
14~TeV (for an integrated luminosity of 10~fb$^{-1}$) or 17~TeV
(for 100~fb$^{-1}$). The reach rapidly deteriorates as $\delta$ is
increased, becoming very sensitive to the cut-off $\Lambda$ and therefore
to the unknown ultraviolet behavior of the theory.

Tree-level graviton exchange is best studied at the LHC in the Drell-Yan
channel, which can probe regions of the $(M_D,\Lambda)$ plane not
accessible to other searches. The case $\delta =1$ allows a prediction
of the rate, which is independent of the ultraviolet cutoff,
with the result that graviton emission is a more sensitive probe than tree-level graviton exchange.

Graviton-loop effects can dominate over tree-level exchange because
they induce operators with lower dimensionality.   
In the regime where gravity can become strong ($\Lambda \gtrsim M_D$),
the search for the effective operator $\Upsilon$ in the Drell-Yan channel
provide the most efficient probe of theories with multi-dimensional
gravity.

\bigskip


\paragraph{Acknowledgments}
We thank G. Isidori, S. Mele and R. Rattazzi for interesting discussions.
We thank I. De Bonis, R. Tenchini and P. Wells for providing and explaining
us the results on $e\bar{e}\to e\bar{e}$ data.

\appendix

\section{Cross sections of graviton processes}\label{cross}
Here we list all cross sections necessary for studying the signals generated by
tree-level exchange of gravitons in $\delta\ge 1$ extra dimensions,
with the following final states:
$f\bar{f}$ and
$\gamma\gamma$ at electron-positron colliders,
and jet jet, $\gamma\gamma$, $\ell\bar{\ell}$ at hadron colliders.
It is useful to define
$$G(s,t)\equiv \frac{s^4 + 10s^3 t+ 42 s^2 t^2 + 64 s t^3 + 32 t^4}{4}.$$
Here $s,t,u$ are the usual Mandelstam variables, related by $s+t+u=0$.
All our cross section are normalized such that the integration range is 
$-s<t<0$, even when 
there are two identical particles in the final state.

The effective graviton propagator ${\cal S}$ in $\delta$ extra dimensions
is defined in eq.~(\ref{spuf}).
For $\delta > 2$ the uncomputable UV divergent contribution dominates and our expressions
can be simplified setting
${\cal S}(s)={\cal S}(t) = {\cal S}(u) ={\cal S}$, which is the limit studied in the previous literature.\footnote{
Our results agree with ref.~\cite{noi},
but we find the following discrepancies with the cross sections for the
hadronic processes computed in ref.~\cite{Soni}.
Their parameter $F/M_S^4$ corresponds to our $-{\cal S}/8\pi$.
In $qq\to qq$, $q\bar{q}\to q\bar{q}$ and $\bar q\bar q\to \bar q\bar q$ scatterings 
our results would agree using instead  $F/M_S^4 \leftrightarrow {\cal S}/4\pi$.
In $gg\to q\bar{q}$ scattering we would agree for  $F/M_S^4 \leftrightarrow- {\cal S}/4\pi$.
The pure graviton contribution to $gg\to gg$ is 4 times bigger than in
ref.~\cite{Soni}.}

\begin{enumerate}
\item {\bf  Electron positron $\rightarrow$ fermion antifermion}.
\begin{eqnarray*}\label{eq:ffbar1}
\frac{d\sigma}{d t}(e^+e^-\leftrightarrow f\bar{f}) &=&
\frac{u^2(|G^{LL}|^2 + |G^{RR}|^2) + 
t^2 (|G^{LR}_s|^2 +|G^{RL}_s|^2) + 
s^2 (|G^{LR}_t|^2 +|G^{RL}_t|^2) }{16\pi s^2}\\
&&+\frac{1}{128\pi s^2}\bigg[|{\cal S}(s)|^2 G(s,t) + |{\cal S}(t)|^2 G(t,s) \\
&&+2\hbox{Re}\,{\cal S}(s)^* \bigg( (G^{LR}_s+G^{RL}_s) t^2 (3s+4t) + (G^{LL} + G^{RR}) (s+4t)u^2\bigg)\\
&&+2\hbox{Re}\,{\cal S}(t)^* \bigg( (G^{LR}_t+G^{RL}_t) s^2 (4s+3t) + (G^{LL} + G^{RR}) (4s+t)u^2\bigg)\delta_{ef}\\
&&+ \frac{\hbox{Re}\,{\cal S}(s) {\cal S}(t)^*}{2} (4s+t)(4t+s)u^2\delta_{ef}\bigg]
\end{eqnarray*}
where $\delta_{ef}=1$ if $f=e$ and $\delta_{ef}=0$ otherwise.
If $f$ is a quark
$d\sigma(e\bar{e}\to f\bar{f})$ must be multiplied by $N_f=3$ while
$d\sigma(f\bar{f} \to e\bar{e})$ must be divided by $N_f$.
The factors 
$$
G^{AB}_s \equiv \sum_{V=\gamma,Z,W} \frac{g_A(V\to \ell\bar{\ell}) g_B(V\to f\bar{f})}{s-M_V^2 + i M_V \Gamma_V} ,  \qquad
G^{AB}_t \equiv \sum_{V=\gamma,Z,W} \frac{g_A(V\to f\bar{\ell}) g_B(V\to \ell\bar{f})}{t-M_V^2} 
$$
($A,B=\{L,R\}$) are the propagators of electroweak SM gauge bosons,
and $G^{LL} = G^{LL}_s + G^{LL}_t$, 
$G^{RR} = G^{RR}_s + G^{RR}_t$.
If $f\neq e,\nu_e$ there is no $t$-channel contribution,  $G_t^{AB}=0$.
In the case of  Bhabha scattering $e^+e^-\to e^+e^-$ one has
$$ 
G^{AB}_x =\frac{e^2}{x}+
 \frac{g_2^2 g_{Ae} g_{Be}/
\cos^2\theta_{\rm W}}{x-M_Z^2+i \Gamma_Z M_Z}\qquad
x = \{s,t\}$$
with $g_{Le} = \sin^2\theta_{\rm W}-1/2$ and $g_{Re} = 
\sin^2\theta_{\rm W}$.\footnote{For 
$\delta = 1$, tree-level virtual graviton effects, 
and in particular the sign of their interference with the
SM, are unambiguously determined.
To compute it correctly it is convenient to focus on the small $t$ limit of Bhabha scattering,
where the cross section does not depend on the spin of the colliding particles
(eikonal approximation). As well known, in such a limit both electromagnetism and gravity
give an attractive force between electrons and positrons, therefore their interference is constructive:
graviton exchange increases the SM $e^+e^-\to e^+e^-$ cross section.

This argument allows to fix related processes, although it 
does not directly apply to $q\bar{q}\to q\bar{q}$ scattering:
gluon exchange and graviton exchange have different colour indices,
which suppress the would-be dominant interference at $|t|\ll s$.}

\item
{\bf  Electron fermion $\to$ electron fermion}
scattering can be obtained from 1.\ by crossing $s\to t$, $t\to u$ and $u\to s$ in the squared amplitude:
\begin{eqnarray*}\label{eq:ffbar2}
\frac{d\sigma}{d t}(e f\to e'f') &=&
\frac{s^2(|G^{LL}|^2 + |G^{RR}|^2) + 
u^2 (|G^{LR}_t|^2 +|G^{RL}_t|^2) + 
t^2 (|G^{LR}_u|^2 +|G^{RL}_u|^2) }{16\pi s^2}\\
&&+\frac{1}{128\pi s^2}\bigg[|{\cal S}(t)|^2 G(t,u) + |{\cal S}(u)|^2 G(u,t) \\
&&+ 2\hbox{Re}{\cal S}(t) ^*\bigg( (G^{LR}_t+G^{RL}_t) u^2 (3t+4u) + (G^{LL} + G^{RR}) (t+4u)s^2\bigg)\\
&&+2\hbox{Re}{\cal S}(u)^* \bigg( (G^{LR}_u+G^{RL}_u) t^2 (4t+3u) + (G^{LL} + G^{RR}) (4t+u)s^2\bigg)\delta_{ef}\\
&&+ \frac{\hbox{Re}\,{\cal S}(t) {\cal S}(u)^*}{2} (4t+u)(4u+t)s^2\delta_{ef}\bigg]
\end{eqnarray*}
Here 
$$
G^{AB}_t \equiv \sum_{V=\gamma,Z,W} \frac{g_A(V\to \ell\bar{\ell}) g_B(V\to f\bar{f})}{t-M_V^2} ,  \qquad
G^{AB}_t \equiv \sum_{V=\gamma,Z,W} \frac{g_A(V\to f\bar{\ell}) g_B(V\to \ell\bar{f})}{u-M_V^2} 
$$
and $G^{LL} = G^{LL}_t + G^{LL}_u$, 
$G^{RR} = G^{RR}_t + G^{RR}_u$.
If $f=e$ $d\sigma(ee\to ee)$ must be divided by 2, if one wants to integrate it
in the full range $-s<t<0$.
At small $t$ SM/graviton interference is destructive in $ee\to ee$ collisions and
constructive in $ep\to ep$ collisions.

\item {\bf  fermion antifermion $\to $ photon photon} 
$$\frac{d\sigma}{dt}(f\bar{f}\to\gamma\gamma) = \frac{t^2+u^2}{64\pi s^2tu N_f}
|2 Q_f^2-tu{\cal S}(s)|^2$$
where $N_f = 1$ and $Q_f = -e$ if $f=e$;
$N_f = 3$ if $f$ is a quark.

\item {\bf  gluon gluon $\to $ photon photon} 
$$\frac{d\sigma}{dt}(gg\to\gamma\gamma) = \frac{t^4+u^4}{512\pi s^2}|{\cal S}(s)|^2$$

\item {\bf  quark antiquark $\to $ quark antiquark}.
Including only the gluon and the graviton contributions we find
\begin{eqnarray*}\label{eq:ffbar3}
\frac{d\sigma}{d t}(q\bar{q}\to q\bar{q}) &=&\frac{1}{16\pi s^2}\bigg[
\frac{8g_3^4}{27s^2 t^2}(s^2-st+t^2)(3s^2+5st+3t^2)\\
&&+\frac{2g_3^2}{9} u^2\hbox{Re}\bigg(\frac{4t+s}{t}{\cal S}(s)^*+\frac{4s+t}{s} {\cal S}(t) ^*\bigg)\\
&&+\frac{|{\cal S}(s)^2| G(s,t) + |{\cal S}(t)^2| G(t,s)}{8} + \frac{\hbox{Re}\,{\cal S}(s) {\cal S}(t)^*}{48}(4s+t)(4t+s)u^2\bigg]
\end{eqnarray*}

\item {\bf  quark quark $\to$ quark quark} is obtained by crossing the previous expression:
\begin{eqnarray*}\label{eq:ffbar4}
\frac{d\sigma}{d t}(q{q}\to q{q}) &=&\frac{1}{32\pi s^2}\bigg[
\frac{8g_3^4}{27 t^2u^2}(t^2-tu+u^2)(3t^2+5tu+3u^2)\\
&&+\frac{2g_3^2}{9} s^2\hbox{Re}\,\bigg(\frac{4u+t}{u} {\cal S}(t)^*+ \frac{4t+u}{t}{\cal S}(u)^*\bigg)\\
&&+\frac{|{\cal S}(t)|^2 G(t,u) + |{\cal S}(u)|^2 G(u,t)}{8} + \frac{\hbox{Re}\, {\cal S}(t) {\cal S}(u)^*}{48}(4t+u)(4u+t)s^2\bigg]
\end{eqnarray*}
and $d\sigma(q{q}\to q{q}) =d\sigma(\bar q\bar {q}\to \bar q\bar {q}) $.

\item {\bf  quark antiquark $\to $ quark$'$ antiquark$'$}.
The prime denotes that the two quarks are of different type.
Including only the gluon and the graviton contributions we find
\begin{eqnarray*}\label{eq:ffbar5}
\frac{d\sigma}{d t}(q\bar{q}\to q'\bar{q}') &=&\frac{1}{16\pi s^2}\bigg[
\frac{4g_3^4}{9s^2}(s^2+2st+2t^2)+\frac{|{\cal S}(s)|^2}{8} G(s,t) \bigg]
\end{eqnarray*}

\item {\bf  quark quark$'$ $\to $ quark quark$'$} and
 {\bf  quark antiquark$'$ $\to $ quark antiquark$'$}.
Including only the gluon and the graviton contributions we find
\begin{eqnarray*}\label{eq:ffbar6}
\frac{d\sigma}{d t}(q\bar{q}'\to q\bar{q}') =\frac{d\sigma}{d t}(q{q}'\to q{q}') &=&\frac{1}{16\pi s^2}\bigg[
\frac{4g_3^4}{9t^2}(t^2+2tu+2u^2)+\frac{|{\cal S}(t)|^2}{8} G(t,u) \bigg]
\end{eqnarray*}

\item {\bf  gluon gluon $\to$ gluon gluon}:
\begin{eqnarray*}\label{eq:ffbar7}
\frac{d\sigma}{d t}(gg\to gg) &=&\frac{1}{256\pi s^2}\bigg[\frac{9g_3^4(s^2+t^2+u^2)^3}{2s^2 t^2 u^2}
\\
&&-6 g_3^2\hbox{Re}\,\bigg( \frac{t^4+u^4}{tu}{\cal S}(s)^* + \frac{s^4+u^4}{su}{\cal S}(t)^*+\frac{s^4+t^4}{st}{\cal S}(u)^*\bigg)\\
&& +u^4(4|{\cal S}(s)|^2+\hbox{Re}\,{{\cal S}(s) {\cal S}(t)^*} +4|{\cal S}(t)|^2) \\
&&+ t^4 (4|{\cal S}(s)|^2+\hbox{Re}\,{{\cal S}(s) {\cal S}(u)^*}+4|{\cal S}(u)|^2)  \\
&&+s^4 (4|{\cal S}(t)|^2+\hbox{Re}\,{{\cal S}(t) {\cal S}(u)^*}+4|{\cal S}(u)|^2)\bigg]
\end{eqnarray*}

\item {\bf  gluon gluon $\leftrightarrow$ quark antiquark}
\begin{eqnarray*}
\frac{d\sigma}{d t}(gg\to q\bar{q}) &=&\frac{1}{16\pi s^2}\bigg[
\frac{-g_3^4}{24s^2t(s+t)}(s^2+2st+2t^2)(4s^2+9st+9t^2)\\
&&-
\frac{g_3^2}{8}(s^2+2st+2t^2)\hbox{Re} {\cal S}(s)^*-\frac{3}{16}|{\cal S}(s)|^2 t(s+t)(s^2+2st+2t^2) \bigg]
\end{eqnarray*}
The cross section for the inverse process is
  $d\sigma(q\bar{q}\to gg )/dt= (32/9)d\sigma(gg\to q\bar{q})/dt$.

\item {\bf  gluon quark $\to $ gluon quark} can be obtained from
the previous result by crossing $s\to t$, $t\to u$. Since we average over the colour  of the initial state,
we must also rescale $\sigma$ by a factor $8/3$. The result is:
\begin{eqnarray*}
\frac{d\sigma}{d t}(gq\to gq) &=&\frac{1}{16\pi s^2}\bigg[
\frac{g_3^4}{9st^2(s+t)}(2s^2+2st+t^2)(9s^2+9st+4t^2)+\\
&&+
\frac{g_3^2 }{3}(2s^2+2st+t^2)\hbox{Re}\, {\cal S}(t) +\frac{1}{2}|{\cal S}(t)|^2 s(s+t)(2s^2+2st+t^2) \bigg]
\end{eqnarray*}
Also,  $d\sigma(gq\to gq )/dt=d\sigma(g\bar{q}\to g\bar{q})/dt$.

\end{enumerate}

Finally we give 
the differential cross sections for the processes $e^+e^-\to \gamma G$
and $e^+e^-\to Z G$:
\beq\frac{d\sigma}{dx\,dy} (e^+ e^- \to V G)= 
\frac{S_{\delta -1}\alpha_V}{M_D^{2+\delta}} s^{\delta/2} y^{\delta/2-1} g_V(x,y,z)\eeq
where $x=t/s$, $y=m^2/s$, $z=M_V^2/s$, $\alpha_\gamma = \alpha_{\rm em}$,
$\alpha_Z =\alpha_{\rm em}(1-4\sin^2\theta_{\rm W} +\sin^4\theta_{\rm W}
)/2\sin^2 2\theta_{\rm W}=0.36 ~\alpha_{\rm em}$, 
and
\begin{eqnarray}
g_Z(x,y,z)&= &
 \frac{1}{96 x^2 
    {(z -1 - x + y   ) }^2}
\left\{
 -4 x^6  ( z-6   )  + 12 x^5  ( z-6   )   ( z-1 + y   )  \right.
  \\   &&- \nonumber
    3 y^2 { ( z-1   ) }^2 z { ( z-1 + y   ) }^2 -     
    2 x^4  \left[ -42 + 6 y^2  ( z-7   )  + 78 z - 58 z^2 + 6 z^3 \right.
      \\   &&+ \nonumber
\left.       y  ( 75 - 114 z + 17 z^2  )   \right]  + 
    3 x y  (z -1   )   \left[ 1 + y^3  ( z-1  )  - 2 z + 10 z^2 - 
10 z^3 + z^4 \right.  
      \\   &&+\nonumber
\left.       
y^2  ( 1 - 14 z + 5 z^2  )  + y  (  13 z-1 - 25 z^2 + 5 z^3  )   \right]  + 
    4 x^3  \left[ 12 + y^3  ( z-12   )  - 25 z \right. 
      \\   &&+\nonumber
\left.      31 z^2 - 19 z^3 + z^4 + y^2  ( 27 - 57 z + 8 z^2  )  + 
       y  ( -27 + 72 z - 69 z^2 + 8 z^3  )   \right]     
       \\   &&+\nonumber
    x^2  \left[ 12 y^4 + y^3  ( -33 + 90 z - 13 z^2  )  
+ 4 { ( z-1   ) }^2  ( 3 + 5 z^2  ) \right.  
     \\   &&-\nonumber \left. \left.
4 y^2  ( -9 + 48 z - 55 z^2 + 8 z^3  )  + y  ( -33 + 102 z - 206 z^2 + 
150 z^3 - 13 z^4  )   \right] \right\}, \\
g_\gamma(x,y)&=&g_Z(x,y,0)=\frac{\left[ 
4x(1+x-y)-y\right] \left[ 1+2x(1+x)-2xy+y^2\right] }{32x(1+x-y)}.
       \end{eqnarray}

\bibliographystyle{plain}\frenchspacing

\begin{multicols}{2}

\end{multicols}

\newpage\normalsize

\centerline{\LARGE\bf \color{red}Erratum}
\vspace{0.3cm}
\bigskip\bigskip

The original version of the paper missed a simple but important difference between flat
and slightly warped extra dimensions, related to the fact that warping reduces the number of Kaluza Klein
gravitons making them more strongly coupled.  In the phenomenologically interesting part of the
parameter space the life-time $1/\Gamma_G \sim \Lambda_\pi^2/m^3$ of gravitons $G$ with mass $m$ is so small that
KK gravitons decay promptly.  This is different from the case of flat extra dimensions, where $\Lambda_\pi \sim M_{\rm Pl}$
and gravitons decay far away from the detector, giving missing energy signals.

As a consequence:
\begin{enumerate}
\item  {\em There are no missing energy signals due to graviton emission}
in the model under consideration.
\item
{\em Graviton production ($i\to G$ followed by $G\to f$ decays) contributes instead
to $i \to f$ scatterings}, where $i$ and $f$ denote initial and final SM particles.
\end{enumerate}
The graviton amplitude ${\cal S}(s)$ was correctly computed here, but the new real contribution
enhances its modulus squared, which becomes:
\beq
\langle |{\cal S}^2| \rangle =  ({\rm Re}\, {\cal S})^2+
\frac{\left( {\rm Im}\, {\cal S} \right)^2}{\epsilon} ,\qquad
\epsilon = \left. \frac{\pi \Gamma_G}{2\Delta m}\right|_{m=\sqrt{s}}=\frac{283}{960}
 \left( \frac{\sqrt{s}}{M_5}\right)^3.
\label{eq:estau33}
\eeq
This enhancement of $|{\cal S}|^2$ is implicit in the equations of~\cite{kiss,kiss2} after averaging it over $\hat{s}$;
the same averaging means that their result for ${\cal S}$ reduces to ours.
The interpretation and consequences of this enhancement were described in~\cite{2011}.

\begin{figure}[h]
$$\includegraphics[width=17cm]{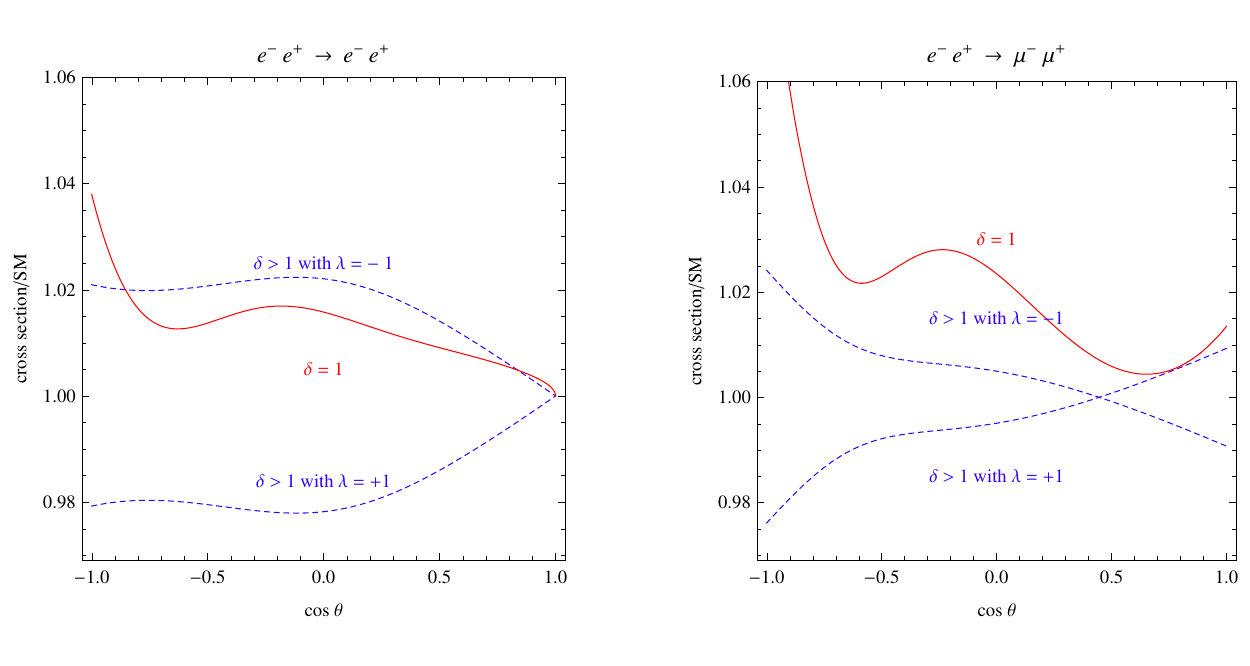}$$\vspace{-1cm}
\caption{\label{fig:LEPvirtualbis}\em Corrections to $e^+e^-\to e^+e^- $ 
(left) and
$e^+e^-\to \mu^+\mu^-$ (right) cross sections
due to tree-level virtual and real exchange of
gravitons with $\delta = 1$ (red continuous line, for $M_D = 2.5\TeV$)
and $\delta >1$ (blue dashed lines, for ${\cal S} = 8\lambda /(1.25\TeV)^4$
and $\lambda=\pm1$).
In the latter case the result depends on the UV cut-off,
so that not even its sign can be reliably computed: we consider
the two cases of constructive and destructive interference with the SM.
For $\delta=1$ the effect is unambiguously fixed: it
increases the  Bhabha cross section, and gives a negligible correction to inelastic scatterings.
}
\end{figure}

In the original version of this work we argued that the interference between the SM and the graviton contribution ${\cal S}$
is significant only for elastic scatterings with  $t$-channel virtual graviton exchange, such as $e^-e^+\to e^-e^+$.  
This remains true, but the
enhanced real graviton amplitude now significantly contributes also to processes
with $s$-channel graviton exchange, such as $e^-e^+\to \ell^-\ell^+$: both the elastic channel ($\ell=e$) 
and the inelastic ones, $\ell=\{\mu,\tau\}$.  This is illustrated in fig.\fig{LEPvirtualbis}, that replaces fig.\fig{LEPvirtual}.

The bound from $e^-e^+\to e^-e^+$ at LEP remains essentially unaltered: $M_D>2.5\TeV$ at $99\%$ confidence level (C.L).
We find a new bound from the total LEP  cross sections into $\mu^-\mu^+$ and $\tau^-\tau^+$:
$M_D>2.4\TeV$ at $99\%$ C.L.
Combining the two bounds we find $M_D>3.3\TeV$ at $95\%$ C.L.
The bound from missing energy signals, such as $e^-e^+\to \gamma G$ does not apply because such signals are not present.

At the LHC, the $pp\to jj$ signal (being dominated by $uu\to uu$ partonic scatterings)
turns out to be negligibly affected by the enhancement in eq.\eq{estau33}~\cite{2011},
which should instead be significant for the $pp\to \ell^-\ell^+$ and $pp\to\gamma\gamma$ signals.

\label{out}

\end{document}